\documentclass[twocolumn]{aastex62}

\usepackage{graphicx}

\usepackage{amsmath}

\shorttitle{Evolution of Cold and Star-Forming Gas in Galaxies}
\shortauthors{Teklu et al.}

\begin{document}

\title{On the Decline of Star Formation during the Evolution of Galaxies}

\correspondingauthor{Adelheid Teklu}
\email{ateklu@usm.lmu.de}

\author{Adelheid Teklu}
\affiliation{Universit\"ats-Sternwarte, Fakult\"at f\"ur Physik, Ludwig-Maximilians-Universit\"at M\"unchen, Scheinerstr.1, 81679 M\"unchen, Germany }
\affiliation{Excellence Cluster Origins, Boltzmannstr. 2, 85748 Garching, Germany}
\author{Rolf-Peter Kudritzki}
\affiliation{Universit\"ats-Sternwarte, Fakult\"at f\"ur Physik, Ludwig-Maximilians-Universit\"at M\"unchen, Scheinerstr.1, 81679 M\"unchen, Germany }
\affiliation{Institute for Astronomy, University of Hawaii at Manoa, 2680 Woodlawn Drive, Honolulu, HI 96822, USA}
\author{Klaus Dolag}
\affiliation{Universit\"ats-Sternwarte, Fakult\"at f\"ur Physik, Ludwig-Maximilians-Universit\"at M\"unchen, Scheinerstr.1, 81679 M\"unchen, Germany }
\affiliation{Max-Planck-Institut f\"ur Astrophysik, Karl-Schwarzschild-Str. 1, 85741 Garching, Germany}
\author{Rhea-Silvia Remus}
\affiliation{Universit\"ats-Sternwarte, Fakult\"at f\"ur Physik, Ludwig-Maximilians-Universit\"at M\"unchen, Scheinerstr.1, 81679 M\"unchen, Germany }\author{Lucas Kimmig}
\affiliation{Universit\"ats-Sternwarte, Fakult\"at f\"ur Physik, Ludwig-Maximilians-Universit\"at M\"unchen, Scheinerstr.1, 81679 M\"unchen, Germany }

\begin{abstract}
Cosmological simulations predict that during the evolution of galaxies, the specific star formation rate continuously decreases. 
In a previous study we showed that generally this is not caused by the galaxies running out of cold gas but rather a decrease in the fraction of gas capable of forming stars.
To investigate the origin of this behavior, we use disk galaxies selected from the cosmological hydrodynamical simulation Magneticum Pathfinder and follow their evolution in time. 
We find that the mean density of the cold gas regions decreases with time.
This is caused by the fact that during the evolution of the galaxies, the star-forming regions move to larger galactic radii, where the gas density is lower. 
This supports the idea of inside-out growth of disk galaxies. 
\end{abstract}



\section{Introduction}

The formation and evolution of galaxies is a complex interplay between different physical effects. 
One of the most important processes involved is the formation of stars from the galaxies' cold gas reservoir. 
In a simplified picture, the galaxies accrete cold gas from the cosmic web, which settles into disks by redistributing angular momentum due to its non-collisionless nature, where then stars are formed. 
During their evolution, star-forming galaxies move along the so-called main sequence, with an increase in stellar mass due to star formation accompanied by a growth in the total amount of gas and thus an increase in star formation rate. However, at a given stellar mass, this star formation rate is generally found to be larger at higher redshifts, with an overall decrease in star formation rate towards lower redshifts \citep[e.g.,][]{santini:2017,Pearson2018}, which leads to a slower growth of the stellar mass. 
When the star formation is shut down (due to several different possible quenching mechanisms), the galaxies fall below the main sequence. 

As the stars form out of molecular gas, understanding the cosmic evolution of the molecular gas mass density is of uttermost importance given that the amount of molecular gas is a natural limiting factor for the amount of possible star formation. Theoretical studies \citep[see e.g.][]{Obreschkow2009a,Obreschkow2009b,Lagos2011,Popping2014} have predicted that from redshift $z\approx2$ to $z = 0$ the cosmic density of the molecular gas decreases.
These predictions were confirmed by several observational studies \citep[see e.g.][]{Walter2014,Decarli2016,Scoville2017,Decarli2019,Riechers2019,Lenkic2020,Tacconi2020}. 
However, measuring the molecular gas masses from observations has its caveats: In most cases, the molecular gas can only be observed through tracers, with the most prominent tracer being CO. Unfortunately, the conversion factor from CO to $H_2$ is still a matter of debate \citep[see e.g.][]{Bolatto2013,Somerville2015}.
Furthermore, \cite{Shetty2014} discuss that CO might not be a direct tracer for the star formation, as it traces the total molecular gas instead of the dense molecular gas \citep{Gao2004}, which is the factual source of star formation. 
Additionally, observational studies found that there is a component of the molecular gas that is actually not star-forming but residing in a thick diffuse disk \citep{Caldu2013,Pety2013}. Therefore, simply assuming the molecular gas mass as tracer of star formation or vice versa may be misleading, and thus connecting the available reservoir in molecular gas to the actual star formation properties is an important yet not well understood issue.

Many modern simulations have shown that with varying the physical parameters, e.g. the modelling of the feedback mechanisms \citep[e.g.][]{Dave2020}, resolution \citep[e.g.][]{Crain2017} or the star formation recipes \citep[e.g.][]{Valentini2019}, the simulations can be fine-tuned to match the observational results. 
This is very helpful for understanding the interplay between the different quantities, but also comes with the caveat that a given property can be obtained through changing different model assumptions while obtaining the same result. Furthermore, comparing simulations directly to observations is not always straight forward. On the one hand, an issue arises when calculating the fractions of $H_2$, as these can vary strongly depending on the used prescription \citep[see e.g.][]{Lagos2015,Valentini2022}. On the other hand, it is important to take selection effects into account when models are compared to observations, as highlighted by \cite{Popping2019}. Additionally, in theoretical works it is commonly assumed that the molecular gas mass is the equivalent of the star-forming gas. This is not only in disagreement with observations as discussed above, but such a non-star-forming $H_2$ component has also been found in simulations \citep{Lagos2015}. Therefore, comparisons between simulations and observations need to take such issues into account to ensure a meaningful comparison.
Nevertheless, to understand the complicated interplay between the available gas reservoir of a galaxy and its resulting star formation is one of the key problems to be deciphered in current galaxy formation studies.

The molecular gas content of a galaxy is thought to be fed from the overall (cold) gas content of a galaxy, and for star forming disk galaxies this reservoir of cold yet not molecular gas in the form of HI is known to be rather extended, in some cases twice the size of the stellar disk, even exhibiting low amount of star formation without showing large molecular clouds \citep[e.g.][]{gildepaz:2005,thilker:2005}. The amount of HI gas around galaxies is known to be larger for galaxies of smaller stellar masses, and flattens to values of about $M_\mathrm{HI} \approx 10^{10}-10^{11}M_\odot$ for galaxies of stellar masses above $M_\mathrm{*} \approx 10^{10}M_\odot$ in the local universe \citep[e.g.,][]{maddox:2015}. However, exploring the gas reservoir of galaxies at higher redshifts is increasingly difficult due to the weakness of the HI 21 cm line, which is the only direct tracer for the HI content \citep[e.g.][]{Chowdhury2022}. However, when comparing to simulations, this total gas content is usually what is the most direct to measure from simulations.

In a previous study by \citet[][hereafter K21]{Kudritzki2021}, we compared look-back models with the hydrodynamical cosmological Magneticum Pathfinder simulations as well as observations, connecting stellar mass, gas mass, and star formation rates in star forming galaxies through cosmic time. We found that during the evolution of galaxies on the main sequence, their star formation rates decline towards low redshifts, not because the galaxies are running out of cold gas, but because the fraction of the star-forming gas declines. This is in excellent agreement with the observed properties as discusses above, however, the reasons for this behaviour are still unclear. In this follow-up study we investigate the physical processes responsible for this decline.

More precisely, we shed light on the process that leads to the decrease of the fraction of the star-forming gas compared to the total cold gas content. For this we use the Magneticum Pathfinder simulations, which are described in Section \ref{sec:sim}. We then show star-forming relations in Section \ref{sec:SFmagneticum} and analyze gas properties in Section \ref{sec:gasprop}. In Section \ref{sec:obs} we compare to observations and conclude our findings in Section \ref{sec:concl}. 


\section{The Magneticum Pathfinder Simulations}\label{sec:sim}

The {\it Magneticum}\footnote{www.magneticum.org} Pathfinder simulations are a set of fully hydrodynamical cosmological
simulations of different box-volumes and resolutions. They follow the formation and evolution of 
cosmological structures through cosmic time, accounting for the complex physical processes which
shape the first building blocks of galaxies into the galaxies seen today. For details on the simulations 
see \citet{Hirschmann2014} and \citet{Teklu2015}. A WMAP-7 $\Lambda\mathrm{CDM}$ cosmology \citep{Komatsu2011}
is adopted with $h=0.704$, $\Omega_m = 0.272$, $\Omega_b = 0.0451$, $\Omega_\lambda = 0.728$, $\sigma_8 = 0.809$,
and an initial slope of the power spectrum of $n_s = 0.963$. 

\subsection{Implementation of Physical Processes}
Star formation and galactic winds are treated in the same way as described by \citet{Springel2003}. 
In this multiphase model for star formation, the inter-stellar medium (ISM) is treated as a two-phase medium, where clouds of cold gas form from cooling of hot gas and are embedded in the hot gas phase assuming pressure equilibrium whenever gas particles are above a given threshold density. 
The density of each gas particle is calculated using a weighted sum over particle neighbors, where the weight decreases with increasing distance.
The hot gas within the multiphase model is heated by supernovae (SNe) and can evaporate the cold clouds. 
Around 10\% of massive stars is assumed to explode as SNe II. 
The released energy by SNe II ($10^{51}$~erg) is modeled to trigger galactic winds with a mass loading rate being proportional to the star formation rate (SFR) to obtain a resulting wind velocity of $v_{\mathrm{wind}} = 350$km/s.
Our simulations also include a detailed model of chemical evolution according to \citet{Tornatore2007}. 

Metals are produced by SNe II, by SNe Ia, and by intermediate- and low-mass stars in the asymptotic giant
branch (AGB). 
Metal radiative cooling rates are implemented according to \citet{Wiersma2009}. 
Metals and energy are released by stars of different mass by integrating the evolution of the stellar population \citep[for details see][]{Dolag2017}, 
properly accounting for mass-dependent lifetimes using a lifetime function according to \citet{Padovani1993},
the metallicity-dependent stellar yields by \citet{Woosley1995} for SNe II, the yields by \citet{Hoek1997} for AGB stars, and the yields by \citet{Thielemann2003} for SNeIa. 
Stars of different mass are initially distributed according to a Chabrier initial mass function \citep{Chabrier2003}.

Our simulations also include a prescription for black hole (BH) growth and feedback from active galactic nuclei (AGN)
based on the model presented by \citet{Springel05a} and \citet{DiMatteo05}, including the same modifications as \citet{Fabjan2010} and some minor changes. The accretion onto BHs and the associated feedback adopt a sub-resolution model \citep[for further details see][]{Hirschmann2014,Steinborn2015,Teklu2015}. 
BHs are represented by collisionless ``sink particles'' that can grow in mass by accreting gas from their environments, or by merging with other BHs. 
They are seeded in galaxies with stellar masses above $\approx 4 \cdot 10^9 h^{-1}M_\odot$ with an initial mass of $\approx 10^5 h^{-1}M_\odot$. 
The gas accretion rate $\dot{M}_\bullet$ is estimated using the Bondi-Hoyle-Lyttleton approximation (\citealp{Hoyle39, Bondi44,Bondi52}): 
\begin{equation}\label{Bondi}
\dot{M}_\bullet = \frac{4 \pi G^2 M_\bullet^2 f_\mathrm{boost} \rho}{(c_s^2 + v^2)^{3/2}},
\end{equation}
where $\rho$ and $c_s$ are the density and the sound speed of the surrounding (ISM) gas, respectively, $f_\mathrm{boost}$ is a boost factor for the density, which typically is set to $100$ and $v$ is the velocity of the BH relative to the surrounding gas. 
The BH accretion is always limited to the Eddington rate (maximum possible accretion for balance between inward-directed gravitational force and outward-directed radiation pressure): 
$\dot{M}_\bullet = \min(\dot{M}_\bullet, \dot{M}_{\mathrm{edd}})$. 
Note that the detailed accretion flows onto the BHs are unresolved, and thus we can only capture BH growth due to the larger-scale gas distribution, which is resolved. 
Once the accretion rate is computed for each BH particle, the mass continuously grows. 
To model the loss of this gas from the gas particles, a stochastic criterion is used to select the surrounding gas particles to be removed. 
Unlike in the model described by \citet{Springel05a}, in which a selected gas particle contributes with all its mass, we included the possibility for a gas particle to lose only a slice of its mass, which corresponds to 1/4 of its original mass. 
In this way, each gas particle can contribute with up to four `generations' of BH accretion events, thus providing a more continuous description of the accretion process.

\begin{figure*}
 \begin{center}
   \includegraphics[width=0.98\textwidth,clip=true]{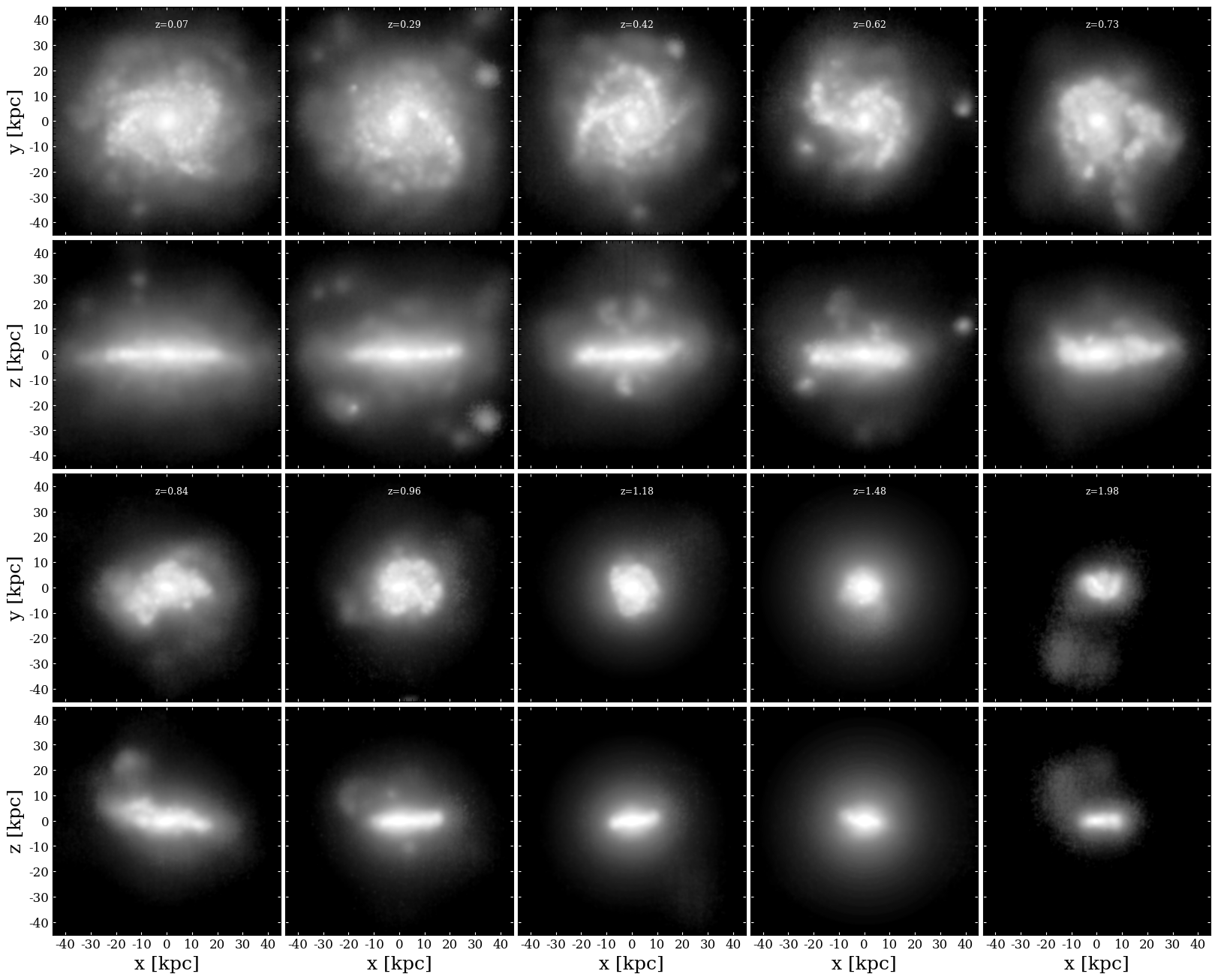}
 \caption{r-Band mock images of an example disk galaxy from the high-mass bin sample, at 10 different redshift bins from $z=0$ (\textit{upper left}) to $z=2$ (\textit{lower right}). First and third row show the face-on view, while second and fourth row show the edge-on views. Mock images were made using the code by \citet{martin:2022}.
 }
  \label{Fig1}
\end{center}
\end{figure*}

\begin{figure*}
 \begin{center}
   \includegraphics[width=0.98\textwidth,clip=true]{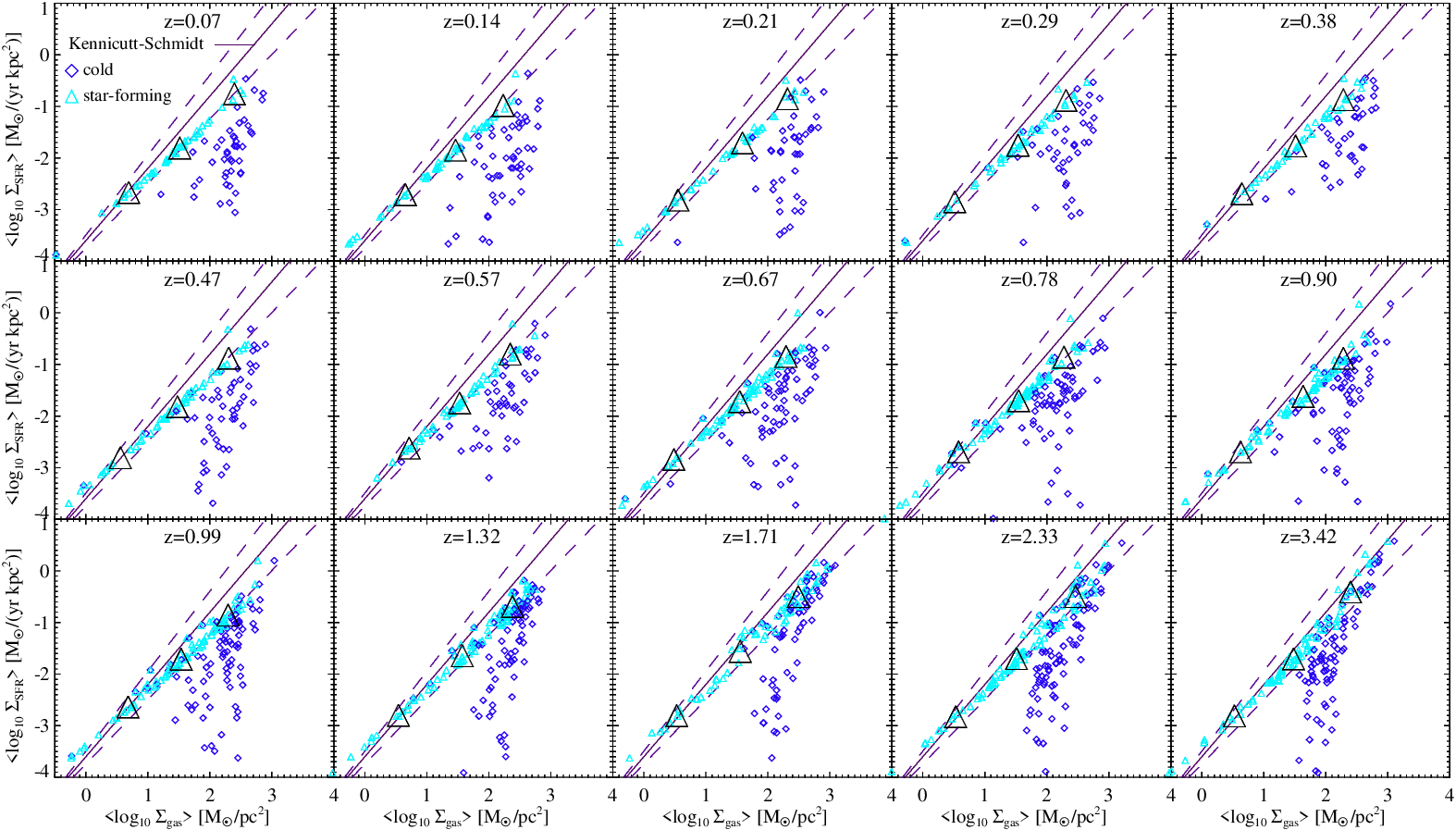}
 \caption{\small{The evolution of the Kennicutt-Schmidt relation for the galaxies in the highest mass bin. The black triangles represent the mean star formation rate density and mean gas surface density within the gas surface density bins, for the star-forming gas. The small data points represent mean values within radial bins of single galaxies, note that one galaxy can have up to 16 data points. Blue diamonds are the cold gas including star-forming gas and turquoise triangles are only star-forming gas. } 
  \label{Fig2} }
\end{center}
\end{figure*}

We assume that a fraction $\epsilon_{\mathrm{f}}$ of the radiated energy is thermally coupled to the surrounding gas so that $\dot{E}_{\mathrm{f}} = \epsilon_{\mathrm{r}} \epsilon_{\mathrm{f}} \dot{M}_\bullet c^2$ is the rate of the energy feedback; $\epsilon_{\mathrm{f}}$ is a free parameter and typically set to $0.1$ \citep[see discussion by][]{Steinborn2015}. The energy is distributed kernel weighted to the surrounding gas particles in an SPH-like manner. Additionally, we incorporated the feedback prescription according to \citet{Fabjan10}: we account for a transition from a quasar- to a radio-mode feedback \citep[see also][]{Sijacki07} whenever the accretion rate falls below an Eddington ratio of $f_{\mathrm{edd}} := \dot{M}_\bullet/ \dot{M}_{\mathrm{edd}} < 10^{-2}$. 
During the radio-mode feedback we assume a 4 times larger feedback efficiency than in the quasar mode. This way, we want to account for massive BHs, which are radiatively inefficient (having low accretion rates), but which are efficient in heating the ICM by inflating hot bubbles in correspondence to the termination of AGN jets. Note that we also, in contrast to \citet{Springel05a}, modify the mass growth of the BH by taking into account the feedback, e.g., $ \Delta M_\bullet = (1-\eta_{r})\dot{M}_\bullet \Delta t$. Additionally, we introduced some technical modifications of the original implementation \citep[for details see][]{Hirschmann2014}.

\subsection{Galaxy Sample Selection}
For our study we use the {\it Magneticum Box4/uhr} simulation, which has a box volume of (48  $h^{-3}$Mpc)$^3$ with initially $2\times576^3$ (dark matter and gas) particles.
The particle masses are $m_\mathrm{DM} = 3.6\times10^7 h^{-1}M_\odot$ and $m_\mathrm{gas} = 7.3\times10^6 h^{-1}M_\odot$, 
respectively, and each gas particle can spawn up to four stellar particles 
(i.e. the stellar particle mass is approximately 1/4th of the gas particle mass), 
with a softening length of $\epsilon_\mathrm{DM}=\epsilon_\mathrm{gas} = 1.4 h^{-1}$ kpc and $\epsilon_*= 0.7 h^{-1}$ kpc.

The analysis of internal evolution of the gas within starforming galaxies requires a high enough resolution. To this end, we select the highest resolution volume simulation, Box4 uhr, which allows for the study of a broad range of stellar masses down to halos of log10(M$_*$/M$_\odot$) = 9.26 which are still resolved with 1000 stellar particles. Previous studies have found galactic properties to agree well with observations \citep[e.g.,][]{Teklu2015,Schulze2018,Remus22}. While the smaller simulation volume may result in missing some rare or particularly unique objects, our focus is on the global trend of galactic evolution for which this volume is sufficient.

For identifying disk galaxies we use their position in the stellar mass--angular momentum ($M_* - j_*$) plane, quantified by the b-value

\begin{equation}
b = \mathrm{log_{10}}\left(\frac{j}{\mathrm{kpc~km/s}}\right) - \frac{2}{3} \log_{10}\left(\frac{M_*}{M_{\odot}}\right),
\end{equation}

where $j$ is the specific angular momentum of the galaxy stellar component (see especially \citealt{Teklu2017}, but also \citealt{Teklu2015} and \citealt{Schulze2018} for more details). 
The connection between morphology and the position on the $M_* - j_*$ plane was first noticed by \cite{Fall1983} and was revisited by \cite{Romanowsky2012}, who proposed scaling relations for disc and elliptical galaxies, which would correspond to large and small b-values, respectively. 
At redshift $z = 0.07$, galaxies with $b \geq -4.35$ are considered disks. 
We select all disk galaxies at $z = 0.07$ with $\log_{10} M_* \geq 9 M_{\odot}$. 
Note that at higher redshift these galaxies could also have been intermediates, spheroids or even passive.

This leaves us with a sample of 621 galaxies selected at $z\approx0$, and for each of these galaxies the evolution is traced back to redshift $z=4.2$. 
The galaxies are divided into 7 bins according to their stellar mass at $z\approx0$, where the mean stellar mass of the galaxies in the bins are $\log_{10} M_*/M_\odot$ = 9.28, 9.65, 10.01, 10.16, 10.34, 10.60, and 11.31, respectively, and the number of galaxies in the according mass bin is 107, 186, 124, 93, 61, 36 and 14. The selected mass bins are an arbitrary choice to represent the range from lower to higher galaxy stellar masses.

Furthermore, we adopt an additional selection criterion for each snap shot based on the SFR law by \citet[][their Appendix C]{Pearson2018} with some modifications, namely with a correction factor as a function of redshift and a modiﬁcation of the power law, which describes the dependence of stellar mass \citep[for details see][]{Kudritzki2021}. We use the threshold of 0.8 dex below our adopted star formation law and include only galaxies above this threshold.

Fig.~\ref{Fig1} depicts r-band mock images of an example disk galaxy from the high mass bin of the Magneticum galaxy sample, followed back in time to show its evolution. For each redshift, the galaxy is shown face-on (upper rows) and edge-on (lower rows). As can clearly be seen, it has a large disk both in stellar component as well as in the gas component (see Fig.~\ref{FigA1} in the Appendix for the gas component) at $z=0$. At higher redshifts, the disk is already present albeit several small merger events had occurred since $z=2$, although as expected the radial extend of the disk is significantly smaller.

In the following, we will study the gas components of the selected disk galaxies in detail, especially the properties of the gas particles. When referring to gas particles that are currently in the process of forming stars, we name those \textit{star-forming gas}. These particles have densities above $10^{7.2} M_\odot \mathrm{kpc}^{-3}$.
If not specified otherwise, the \textit{cold gas} component includes gas particles with temperatures below $10^5 K$ and star-forming gas. The \textit{hot gas} component consists of gas particles which have temperatures above that temperature threshold and are non-star-forming.


\section{Star Formation in Magneticum}\label{sec:SFmagneticum}

 As described previously, star formation is treated within a sub-resolution model \citep[][]{Springel2003}, where gas above a certain density threshold is treated as a two-phase medium. Above this threshold, cold, star-forming clouds form from cooling of hot gas and are embedded in the hot gas phase assuming pressure equilibrium plus a stellar component. 
This sub-resolution model describes the unresolved ISM by solving a set of connected differential equations for mass and energy flows with a closure condition described in detail by \citet[][and references therein]{Springel2003}. 
Within this sub-resolution model, every gas element in the simulation has an associated star formation rate following from the solution of this set of differential equations. This allows then to map in detail the star formation within the simulated galaxies.

\begin{figure*}
 \begin{center}
   \includegraphics[width=0.95\textwidth,clip=true]{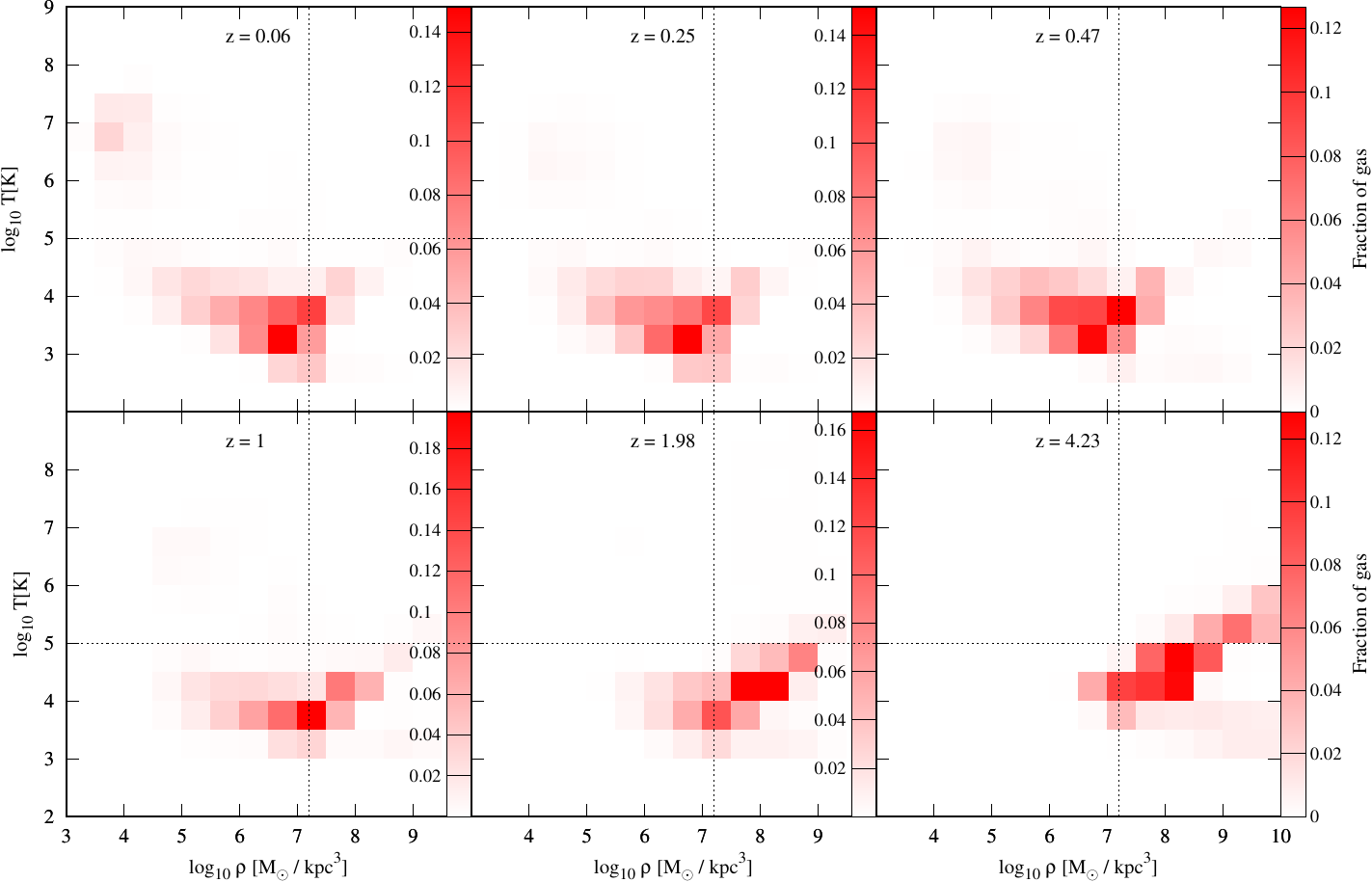}
 \caption{\small{The evolution of the mean temperature-density diagram for the gas of the galaxies in the highest mass bin. The intensity of the red color encodes the fraction of gas resolution elements in a grid cell. The dashed vertical line represents the star formation threshold, and the horizontal line the temperature cut for cold gas. For examples of different mass bins, see Fig. \ref{FigA3} in the Appendix. Note that we chose to show only six representative timesteps in this figure different than in other figures depicting time evolution in this study, as in these mean stacked evolution plots the overall trend is important to see and individual smaller timesteps in this case do not differ strongly but rather demonstrate a smooth trend as can clearly be seen from the six timesteps shown here.
 Also note that the color bar changes according to the maximum fraction of the shown timestep.}
  \label{Fig3} }
\end{center}
\end{figure*}
First and foremost, the question is how the gas reservoir of a given galaxy evolves with redshift with respect to its global star formation properties. Thus, in Fig. \ref{Fig2} we show the mean gas surface density against the mean star formation rate surface density, i.e. the Kennicutt-Schmidt (KS) relation, of the cold gas (including star-forming gas, dark blue diamonds) and only the star-forming gas (turquoise triangles), for the galaxies in the highest mass bin as an example. 
Each data point represents the mean value of one galaxy in a radial bin out to $40\mathrm{kpc}$ with 16 equally spaced bins à $2.5\mathrm{kpc}$. 
Black triangles show the average star formation rate density and average gas surface density within gas surface density bins, for the star-forming gas.
The star-forming gas of the galaxies, on average, follows the evolution according to the KS-law \citep[][solid lines]{Kennicutt1998} within its uncertainties (dashed lines), at all redshifts up to $z\approx3.5$.
We find that at high redshifts the galaxies exhibit higher densities both in the cold gas but also in star formation rate, and that these densities decrease towards low redshifts. 
The cold gas falls off the relation toward lower SFR-densities as in these densities the mean values are dominated by cold non-star-forming gas. 
This is in agreement with previous observational studies, for example \cite{Bigiel2008} found that the relationship of the total gas surface density and the SFR surface density varies within and between the galaxies, and that there is almost no correlation between the surface density of the HI gas, which is generally associated with the cold non-star-forming gas, and the SFR surface density. 
Interestingly, this deviation of the cold gas from the KS-law is relatively constant over time. 
This points out the importance of using only the star-forming gas instead of the cold gas, when investigating and comparing the gas properties to observations. 
Note that, as the star-forming gas is a subsample of the cold gas, the values for the SFR surface density are the same for each radial bin, while the gas surface density is equal or lower for the star-forming gas compared to the cold gas. 
In Fig. \ref{FigA2} in the Appendix we show the KS-relation for all selected galaxies at $z=0.06$ in order to demonstrate that this relation is valid for all mass bins. 

In Fig. \ref{Fig3} (see also Fig. \ref{FigA3} in the Appendix) we show the evolution of the stacked temperature--density phase diagrams for all gas of galaxies in the highest mass bin, where the intensity of the red color mirrors the fraction of the resolution elements of gas in the grid cell. 
The dotted vertical line represents the density threshold of 7.2 $M_\odot kpc^{-3}$, above which the star formation sets in, and the dotted horizontal line shows the temperature cut made to distinguish between cold and hot gas. 
At high redshift most of the gas resolution elements are cold and dense and thus have a high star formation activity.  
Towards lower redshift there is still a large amount of cold gas, which however becomes less dense and thus does not reach the star-forming threshold anymore. 
At the lowest redshift we can see the build-up of a hot and diffuse gas component, i.e. gas with high temperature and low density, which corresponds to gas that has been heated, for example by AGN feedback.

\begin{figure*}
 \begin{center}
   \includegraphics[width=0.98\textwidth,clip=true]{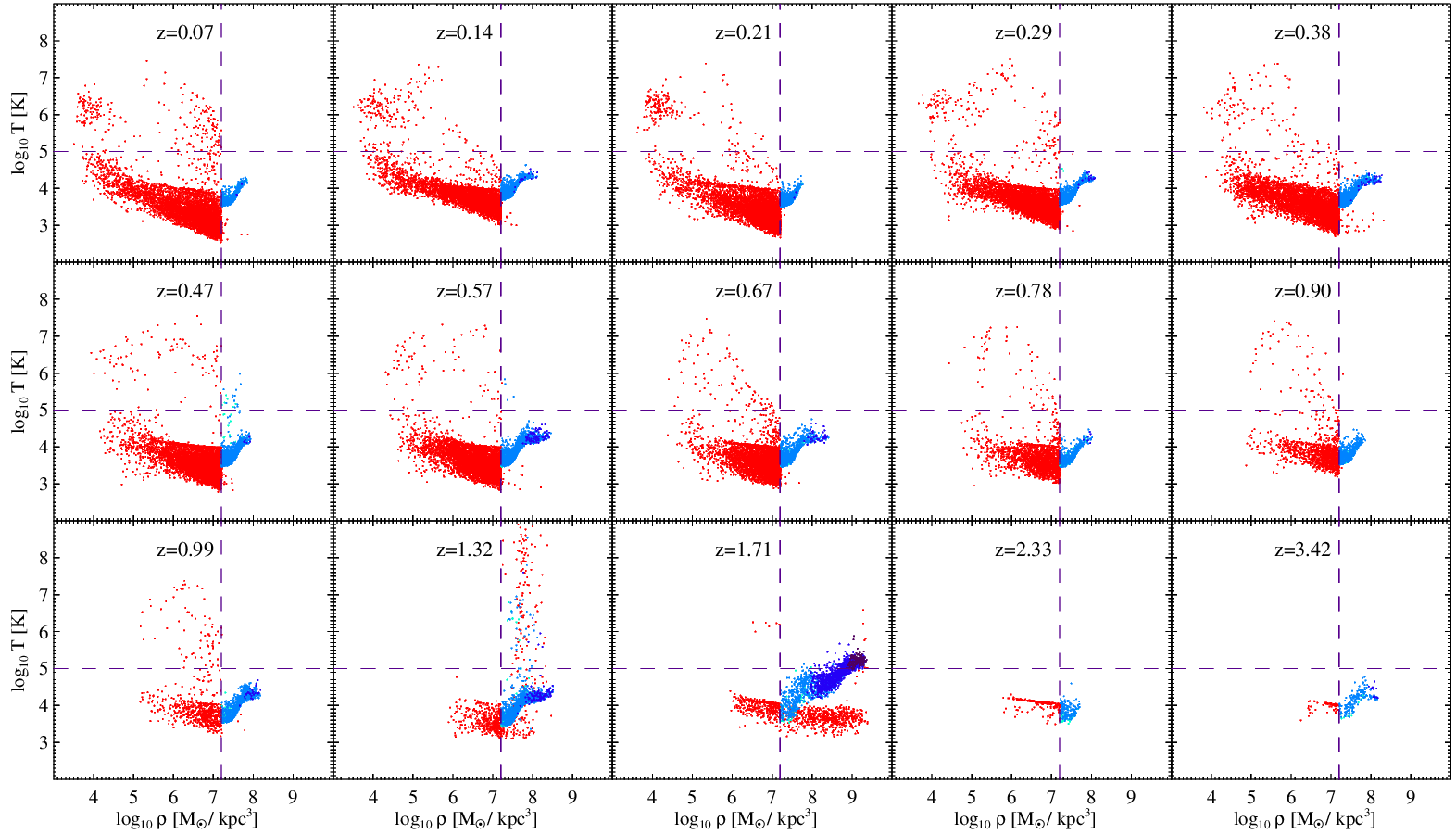}
 \caption{\small{The evolution of the temperature-density diagram for one example galaxy, where each data point represents the physical properties within a resolution element. Red is gas without any star-forming component, star-forming gas ranges from turquoise and blue tones to dark blue (with increasing activity). The dashed vertical line represents the star formation threshold, and the horizontal line the temperature cut for cold gas.}
  \label{Fig4} }
\end{center}
\end{figure*}

As in the stacked phase diagram possibly interesting signatures could get lost, we show in Fig. \ref{Fig4} the evolution of the phase diagram for one example galaxy from the highest mass bin, the same as shown in Fig.~\ref{Fig1}. 
Each data point represents one gas resolution element of the galaxy within $5 R_{1/2}$. 
The colors indicate their star-forming activity, where red is non-star-forming, while turquoise, the different blue tones and dark blue are star-forming (with increasing activity).
At high redshifts most of the gas is cold and dense, and forms stars. This changes with decreasing redshift, where the amount of cold gas which is not dense enough to reach the SF threshold increases and star formation decreases, respectively. Still, at low redshift there is a large amount of cold gas. Note that gas which is assumed to be in the wind phase, is prevented from forming stars and therefore some gas even when above the SF threshold does not form stars.


\section{Evolution of Cold and Star-Forming Gas}\label{sec:gasprop}

In the following, we will study different quantities of the cold and star-forming gas, such as the mass, the density and the radial profiles in order to shed light on the origin of the decrease of the star-forming gas fraction. 
\begin{figure}
 \begin{center}
   \includegraphics[width=0.4\textwidth,clip=true]{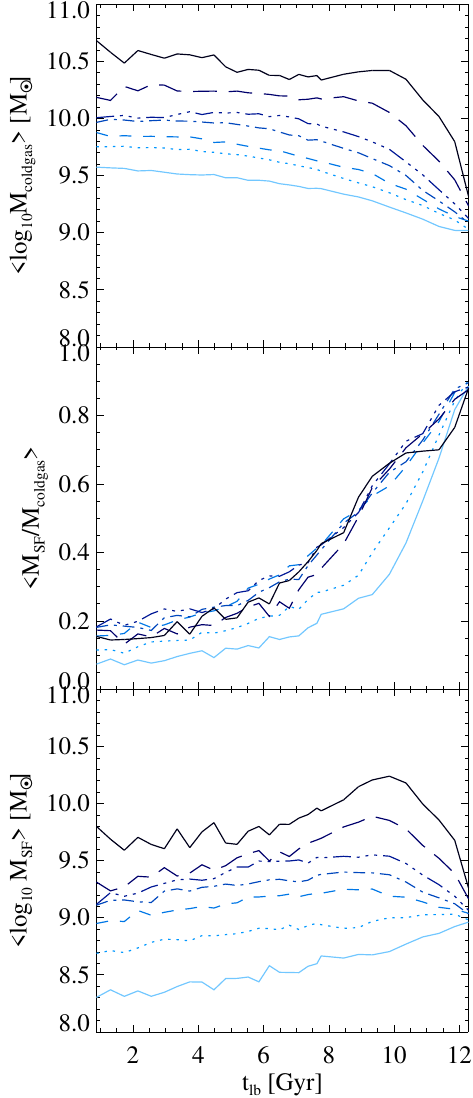}
 \caption{\small{The colors and lines encode the mean values of the galaxies in the different stellar mass bins, with mean $log_{10} M_*/M_\mathrm{sol}$ values of 9.3, 9.66, 9.95, 10.14, 10.33, 10.59 and 11.15 (from light blue representing the lowest mass bin to dark blue showing the most massive bin). 
Top: The cold gas mass increases with time for the three lowest mass bins, whereas for the higher mass bins it remains relatively constant after the first steep increase. 
Middle: The fraction of the star-forming gas mass compared to the cold gas content decreases continuously, where there is no difference between the mass bins except for the two lowest ones. 
Bottom: Initially, the mass of the star-forming gas increases for the high mass bins and then decreases, while it decreases for the low mass bins.} 
  \label{Fig5} }
\end{center}
\end{figure}
Fig. \ref{Fig5} shows the time evolution of the cold gas mass (top panel), the fraction of the star-forming gas compared to the cold gas (middle panel), and the star-forming gas mass (bottom panel). 
For this, we calculate the mean values of the galaxies in the different stellar mass bins, which are encoded by different shades of blue with dark blue being the highest mass bin, the next lower mass bins becoming lighter, and light blue being the lowest mass bin. 
As we focus on the global trends and not those of individual galaxies, and as it would be confusing to show the lines of the errors or the scatter, in Fig. \ref{FigA4} in the Appendix we show the standard deviation exemplary. 
The mean cold gas mass continuously increases with time for the three lowest mass bins, whereas for the higher mass bins it remains relatively constant after the first rapid increase. 
We can clearly see that the galaxies do not run out of cold gas. 
Instead, as already shown in K21, the fraction of the star-forming gas decreases with evolving time independently of the stellar mass, as aside from the two lowest stellar mass bins the curves generally lie on top of each other. 
Interestingly, at early times the fraction of the star-forming gas is very similar for all mass bins; however, the evolution then differs, as the mass of the star-forming gas increases and then decreases for the galaxies in the high mass bins, while for those in the two lowest mass bins it decreases from the beginning. 
This highlights that it is not the lack of cold gas for galaxies in any of the stellar mass bins but the capability of the gas to form stars, which leads to a decrease of the star formation.
We note that, given the smooth trends found between the various mass bins in mean star-forming and cold gas content, we expect that a variation of the mass bin borders would result in little change of the findings presented here. However, the lowest mass bin is to be taken with caution as these galaxies are very small at all redshifts.

\begin{figure}
 \begin{center}
   \includegraphics[width=0.45\textwidth,clip=true]{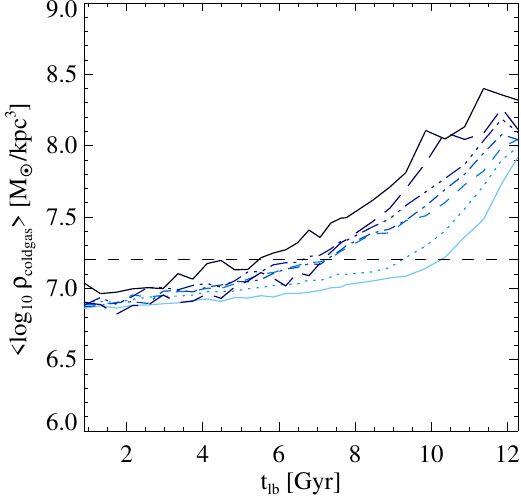}
 \caption{The mean logarithmic mass density of cold gas as function of look-back time, where the different mass bins are color-coded as in Fig. \ref{Fig5}. 
 The black dashed line represents the threshold for star formation. 
\label{Fig6} }
\end{center}
\end{figure}
\begin{figure}
 \begin{center}
   \includegraphics[width=0.45\textwidth,clip=true]{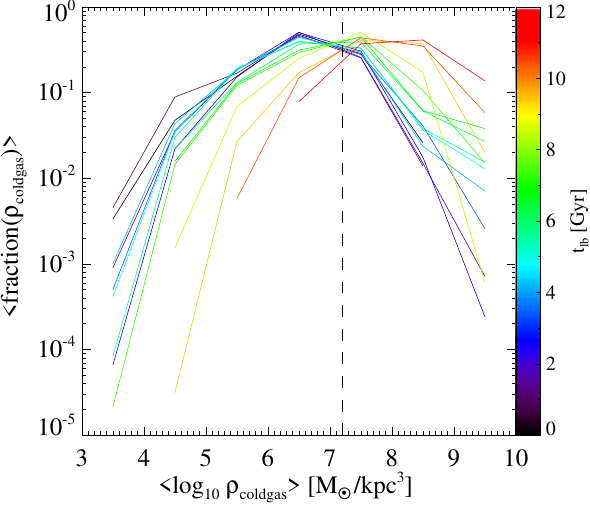}
 \caption{The fraction of cold gas resolution elements with a given density for the highest mass bin, where the different colors and lines represent the different look-back times. 
 The dashed line represents the threshold for star formation. Towards lower look-back times the peak shifts from high to low density, lying below the SF threshold at later times. 
\label{Fig7} }
\end{center}
\end{figure}
\begin{figure*}
 \begin{center}
   \includegraphics[width=0.43\textwidth,clip=true]{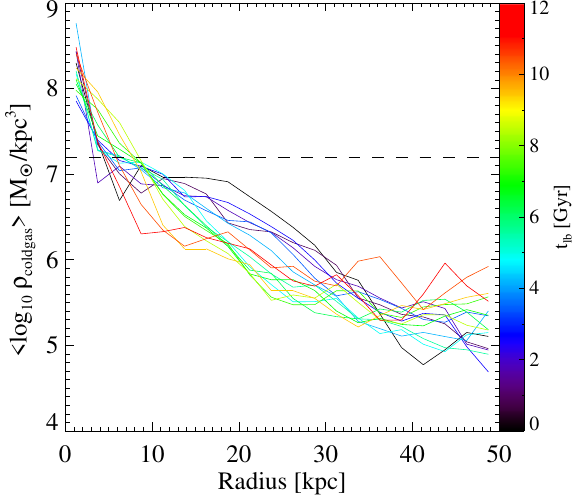}
   \includegraphics[width=0.43\textwidth,clip=true]{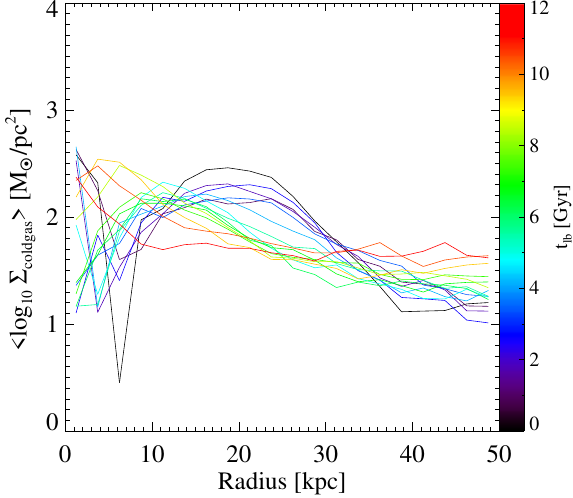}
 \caption{Left: The radial profile of the mean logarithmic mass density of cold gas for the highest mass bin, where the different colors and lines represent the different look-back times and the dashed line represents the threshold for star formation. Right: The radial profile of the mean logarithmic surface density of cold gas. 
\label{Fig8} }
\end{center}
\end{figure*}

As the stars form out of cold gas which condenses to dense clouds, we next study the density of the cold gas in Fig. \ref{Fig6}. 
Here, we use the (mass) density of the cold gas, which is calculated using a weighting function around the resolution elements and in this way roughly expresses the local densities, i.e. these gas resolution elements represent large gas regions in which stars can form. 
This is in contrast to averaging over parts of the galaxy or the whole galaxy, as e.g. done for the cosmic gas density. 
In Fig. \ref{FigA5} of the Appendix we show that the ``global'' density $= M_\mathrm{coldgas} / (4/3\cdot \pi \cdot (5 \cdot R_{1/2})^3)$ results in different curves.
In Fig. \ref{Fig6} we find that with evolving time the mean density of the cold gas decreases for all galaxies. 
It then drops below the threshold for the star formation (black dashed line), which happens earlier for low mass galaxies and later for high mass galaxies.

In the following we want to explore why this mean density of the cold gas decreases with time. 
For this, we analyze the galaxies in the highest mass bin in more detail.  
Fig. \ref{Fig7} shows the average fraction of cold gas in $5 \cdot R_{1/2}$ within a certain density bin. 
Here, the different lines and colors encode the look-back times and the dashed line the density threshold for star formation. 
As clearly can be seen, at high redshift most of the cold gas has a density above the star formation threshold. 
Towards lower redshifts the distribution becomes broader and the peak is shifting to lower density, below the star formation threshold.
So while the total amount of cold gas at low redshifts is still rather large, its density clearly is too low for most of it to continue star formation. This raises the question what is causing this behavior.

To understand this, we take a closer look at the radial distribution of the cold and star-forming gas, as this aspect has so far been neglected in the above quantities. 
The left panel of Fig. \ref{Fig8} shows the radial profile of the average gas density in a certain radial bin. 
The different lines and colors represent the different look-back times and the dashed line represents the density threshold for star formation. 
Overall, the mean density decreases with evolving time. 
Another general, redshift independent trend is that the density is higher in the center and decreases towards larger radii. 
This can be explained by the hydrostatic pressure equilibrium, where the gas experiences the gravitational force and thus pressure and density are higher in the center, where most of the galaxy's mass resides.
We note that in the center of the galaxies, especially at small look-back times, some of the galaxies have a ring of cold gas and thus it is possible that only a small number of galaxies contributes to the data points in the  very central regions. 
However, this plot does not account for the number of gas resolution elements nor the mass of the gas that is found in these radial bins.  
Therefore, on the right panel of Fig. \ref{Fig8} we show the surface density of the cold gas, as this quantity takes the spatial distribution better into account. 

Interestingly, we clearly see that the maximum of the surface density is moving to larger radii towards low redshifts. 
Note that for all radial bins we only consider galaxies with some amount of cold gas to calculate the mean surface density. As many galaxies in our sample at the lowest redshifts have little or no cold gas at their centers given that many of them have rings of cold gas as mentioned already above, the mean values for the two centralmost radial bins are dominated by the few galaxies which do, resulting in a spurious peak from the low number statistics for these two bins. The clear peak in cold gas density moving towards larger radii with lower redshifts, however, is not affected by low number statistics and is a real result of physical processes.

\begin{figure}
 \begin{center}
   \includegraphics[width=0.42\textwidth,clip=true]{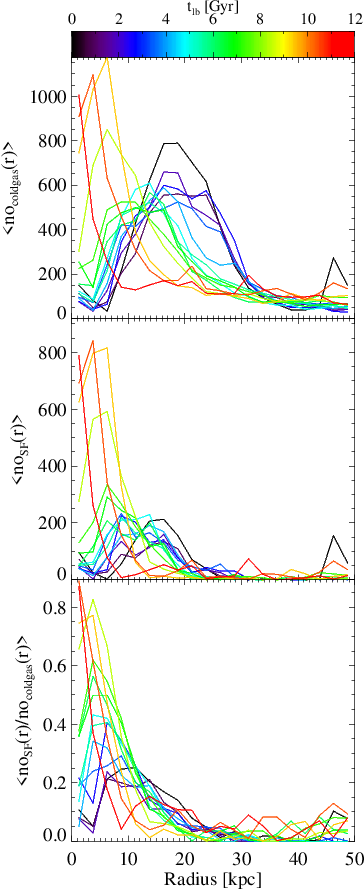}
 \caption{The mean number of cold gas (top), star-forming (middle) resolution elements and the fraction of star-forming to cold gas resolution elements (bottom) in each radial bin for the galaxies in the highest mass bin, where the different colors and lines represent the different look-back times.
\label{Fig9} }
\end{center}
\end{figure}

Fig. \ref{Fig9} shows the mean number of gas resolution elements in a certain radial bin, with the top panel showing the cold gas, the middle panel showing the star-forming gas and the bottom panel the fraction of star-forming to cold gas (including star-forming gas). 
We can clearly see that at large look-back times the majority of the gas is in the center and that the peak moves further outside with evolving time. 
The distribution seems to become broader, which reflects that the total number of cold gas resolution elements does not decrease. 
However, the peak of the star-forming gas becomes noticeably smaller, albeit it also moves to larger radii. 
The fraction shown in the bottom panel displays that the star-forming gas is becoming less over time as the cold gas regions move to larger radii. 
To further illustrate that the star-forming regions are moving outside with decreasing look-back times, in Fig. \ref{FigA1} in the Appendix we show the evolution of one example galaxy projected to the $xy$- and $xz$-plane. 


\section{Comparison with Observations}\label{sec:obs}

For the comparison with observations we use the star-forming gas as a proxy for the H$_2$ gas. 
However, this has to be taken with some caution, as e.g. \cite{Lagos2015} have found that in the EAGLE simulations this approximation is poor for H$_2$ content above a certain mass and that the H$_2$ masses are depending on the resolution.  
As a proxy for the atomic hydrdogen we use the cold gas excluding the star-forming gas. 
We calculate the gas and stellar masses within a radius of 40 kpc in order to cover a similar size compared to the observational results by \cite{Chowdhury2022}, who analyzed cubes of 90 kpc length for their samples of star-forming galaxies to investigate the HI contents at high $z$. 

\begin{figure}
 \begin{center}
   \includegraphics[width=0.45\textwidth,clip=true]{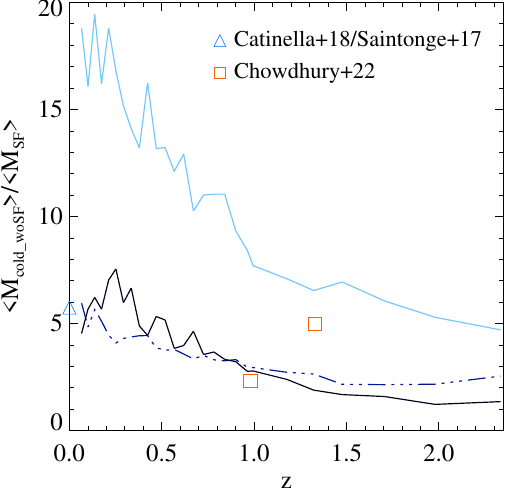}
 \caption{\small{The redshift evolution of the mean mass of the cold gas without SF gas divided by the mean mass of the SF gas only. The observational data (red squares) for $z\approx 1$ and $z\approx 1.3$ were presented in \cite{Chowdhury2022} and the estimated observational value for $M_\mathrm{atomic}$ and $M_\mathrm{molecular}$ at $z\approx0$ were taken from \cite{Catinella2018} and \cite{Saintonge2017}, respectively. The colors encode the lowest (light blue) and highest (dark blue) as well as the mass bin corresponding approximately to the mass the observations covered (blue dash-dotted line). }  
\label{Fig10} }
\end{center}
\end{figure}

Fig. \ref{Fig10} shows the mean mass of the cold gas without star-forming gas divided by the mean mass of the star-forming gas, and the observed mean mass of atomic gas divided by the mean molecular gas mass, respectively. For better visibility, we only show three mass bins, namely the highest mass bin (dark blue), the lowest mass bin (light blue) and the mass bin of about $10^{10} M_\odot$ (blue dash-dotted line) roughly corresponding to the mean stellar mass used in the observations. 
The fractions decrease with increasing redshift for the galaxies in our simulation. 
They agree well with the observations for $z\approx 0$ and $z\approx1$. 
On the other hand, we find some disagreement with the \cite{Chowdhury2022} value at $z\approx1.3$, which indicates an increasing fraction. We note, however, that \cite{Chowdhury2022} discuss potential uncertainties for the mass of molecular gas, which would lower their fraction to 2.5.

Our results are in agreement with the trend presented by \cite{Obreschkow2009b}, and \cite{Power2010}, who analyzed predictions of semi-analytic models by \cite{Baugh2005,Bower2006,Delucia2007,Font2008}, where the fraction of the molecular compared to atomic hydrogen is increasing towards higher redshifts. 
Similarly, \cite{Lagos2011} using a galaxy formation model find that the fraction of H2 compared to HI increases to a peak at $z\approx3.5$, then H2 dominates until $z\approx5$ and above this redshift HI is the dominant component of the ISM. 


\section{Summary and Discussion}\label{sec:concl}

We have analyzed a sample of galaxies in the cosmological hydrodynamical simulation Magneticum Pathfinder. 
The galaxies were selected at $z = 0.1$ such that they are disk galaxies according to the $b$-value. 
These galaxies were put into mass bins and were followed backwards in time.
In this way, we calculate average properties for different mass bins and study the average evolution of today's galaxies that are similar in mass at present instead of averaging over a population at different points in time.
Our aim was to shed light on the question why the fraction of star-forming gas is decreasing, while there is still much cold gas in these galaxies at low redshift \citep[see also][]{Kudritzki2021}. 
Our analysis has shown the following:

\begin{itemize}
\item The density distribution of cold gas clouds is shifting from higher to lower densities with evolving time. As a result, the number of clouds reaching the threshold for star formation is decreasing. 
\item The reason for this behavior is a shift of the maximum of the number distribution of clouds to larger galactic radii indicating an inside-out growth of disk galaxies, in agreement with observations \citep[see e.g.][]{Gonzalez2015,Goddard2017}. Since the gas clouds are in pressure equilibrium with the galaxy's gravitational potential, their density is smaller at larger radii and a larger fraction of their distribution falls below the star formation threshold. 
\end{itemize}

While the average density of the cold gas in the whole galaxy as well as seen in the outer parts of the galaxies is below the star formation threshold (especially towards low redshifts), individual regions of cold gas still reach densities high enough to form stars. These regions are shifting towards larger radii with evolving time. 

In summary, the fraction of the star-forming gas is simply becoming smaller as the gas clouds move further out to larger radii during the evolution of the galaxy, where the gas density is smaller. 

Our findings show the need to better consider not just the total galactic gas content, but also where it is radially distributed. The inside-out growth of disk galaxies provides a natural mechanism for a soft quenching, not for lack of cold gas material but rather due to lack of pressure. Processes which clump together this gas may then result in reignition of the star formation.


\acknowledgments
We thank Andi Burkert and Tadziu Hoffmann for useful discussions. 
We thank our referee for engaged and very constructive comments.
This work has been supported by the Munich Excellence Cluster Origins
funded by the Deutsche Forschungsgemeinschaft (DFG, German Research
Foundation) under Germany's Excellence Strategy EXC-2094
390783311. KD and LK acknowledge support by the COMPLEX project from 
the European Research Council (ERC) under the European Union’s Horizon 
2020 research and innovation program grant agreement ERC-2019-AdG 882679.
The Magneticum Pathfinder simulations were
performed at the Leibniz-Rechenzentrum with CPU time assigned to
the Project ``pr86re''. We are especially grateful for the support by M.
Petkova through the Computational Center for Particle and Astrophysics
(C2PAP). 

\appendix


\section{Spatial Distribution of Gas and Stars}

\begin{figure*}
 \begin{center}
   \includegraphics[width=0.8\textwidth]{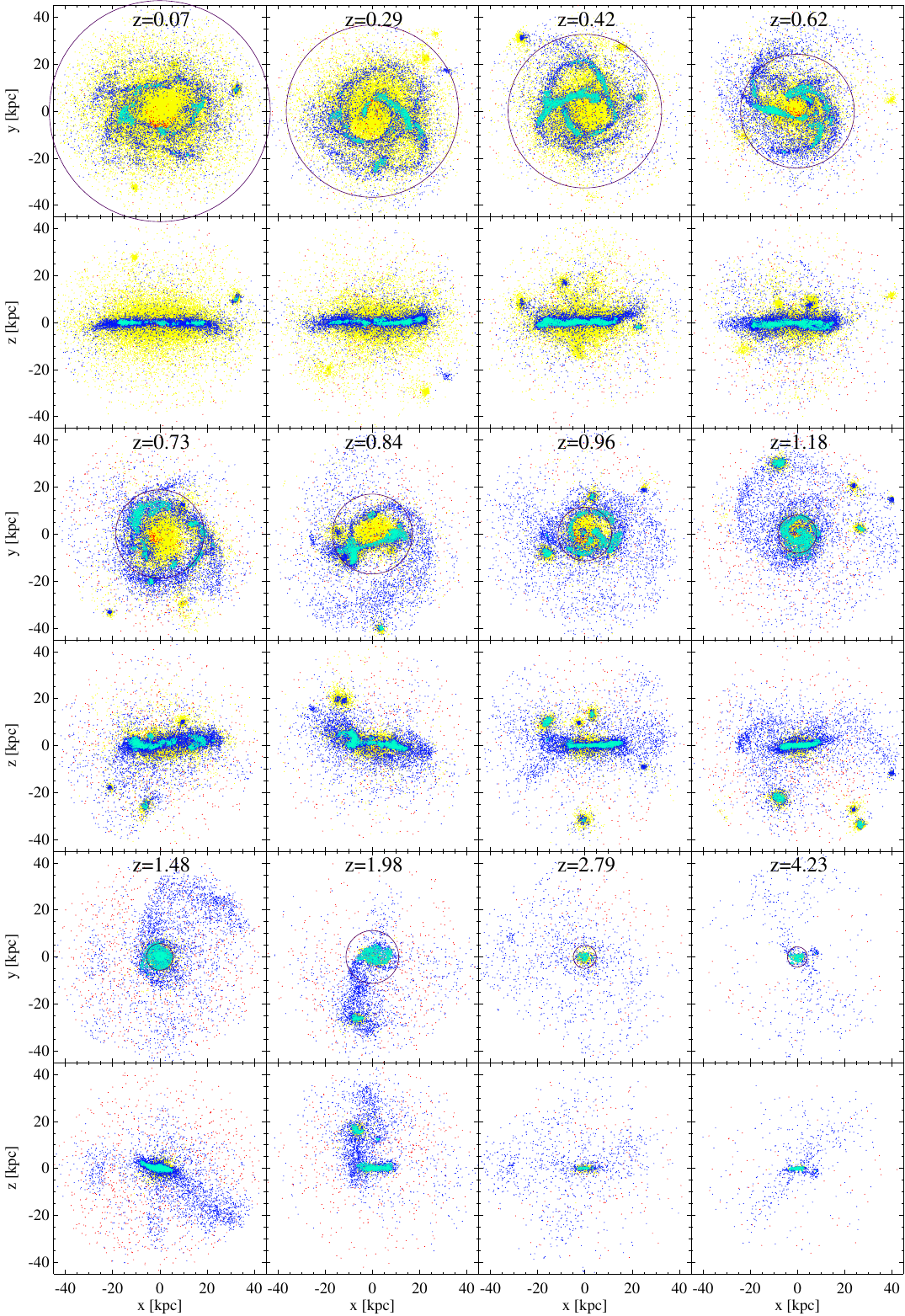}
 \caption{\small{The evolution of the spatial distribution of stars (yellow), hot gas (red), cold gas (blue) and star-forming gas (turquoise) for one example galaxy. The circle indicates $5 \cdot R_{1/2}$.}
\label{FigA1} }
\end{center}
\end{figure*}

Fig. \ref{FigA1} shows the evolution of an example galaxy in the highest mass bin. For twelve redshifts we plot the spatial distribution of stars (yellow), hot gas (red), cold gas (blue) and star-forming gas (turquoise) in face-on projection (upper panels, odd rows) and edge-on projection (lower panels, even rows). The circle indicates $5 \cdot R_{1/2}$. 
This figure illustrates that at early times the galaxy is small and compact and that the star formation takes place in the inner part. With evolving time the gas is pushed outwards and the stars form at larger radii. 

\section{Kennicutt-Schmidt Relation for All Disk Galaxies at $z=0$}

\begin{figure*}
 \begin{center}
   \includegraphics[width=0.7\textwidth]{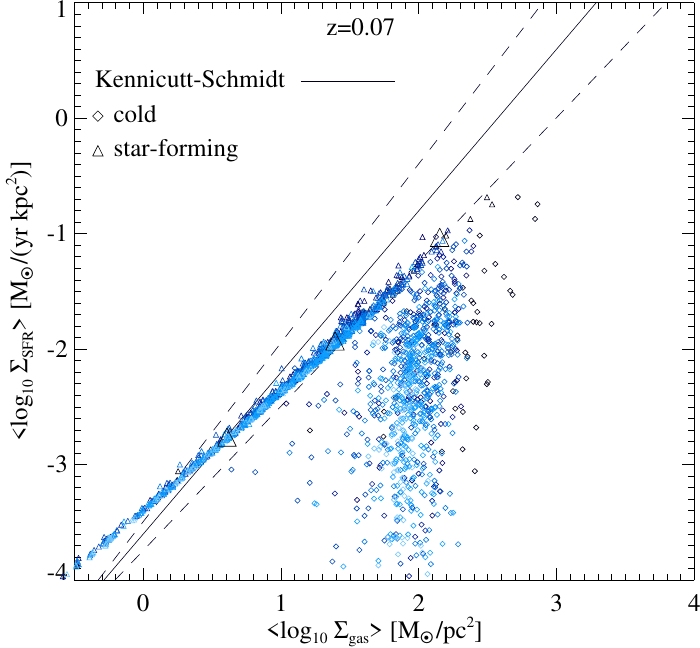}
 \caption{\small{The KS relation for all selected star-forming disk galaxies.}
\label{FigA2} }
\end{center}
\end{figure*}

Fig. \ref{FigA2} shows the Kennicutt-Schmidt relation for all galaxies fullfilling the selection criteria at redshift $z=0$.  The colors encode the mass bins of the galaxies, where light colors represent the low mass galaxies and dark blue the high mass galaxies. 
We note that we omitted the innermost radial bin due to poor resolution.
As discussed above, our results are in agreement with previous observational studies. However, we only consider late-type galaxies with a certain amount of cold and star-forming gas. So with our sample of galaxies, we cannot predict whether the low surface brightness galaxies in our simulation fall below the KS relation as found in observational studies \citep{Bigiel2008,Nagesh2023}.

\section{Temperature vs. Density}

\begin{figure*}
 \begin{center}
   \includegraphics[width=0.9\textwidth,clip=true]{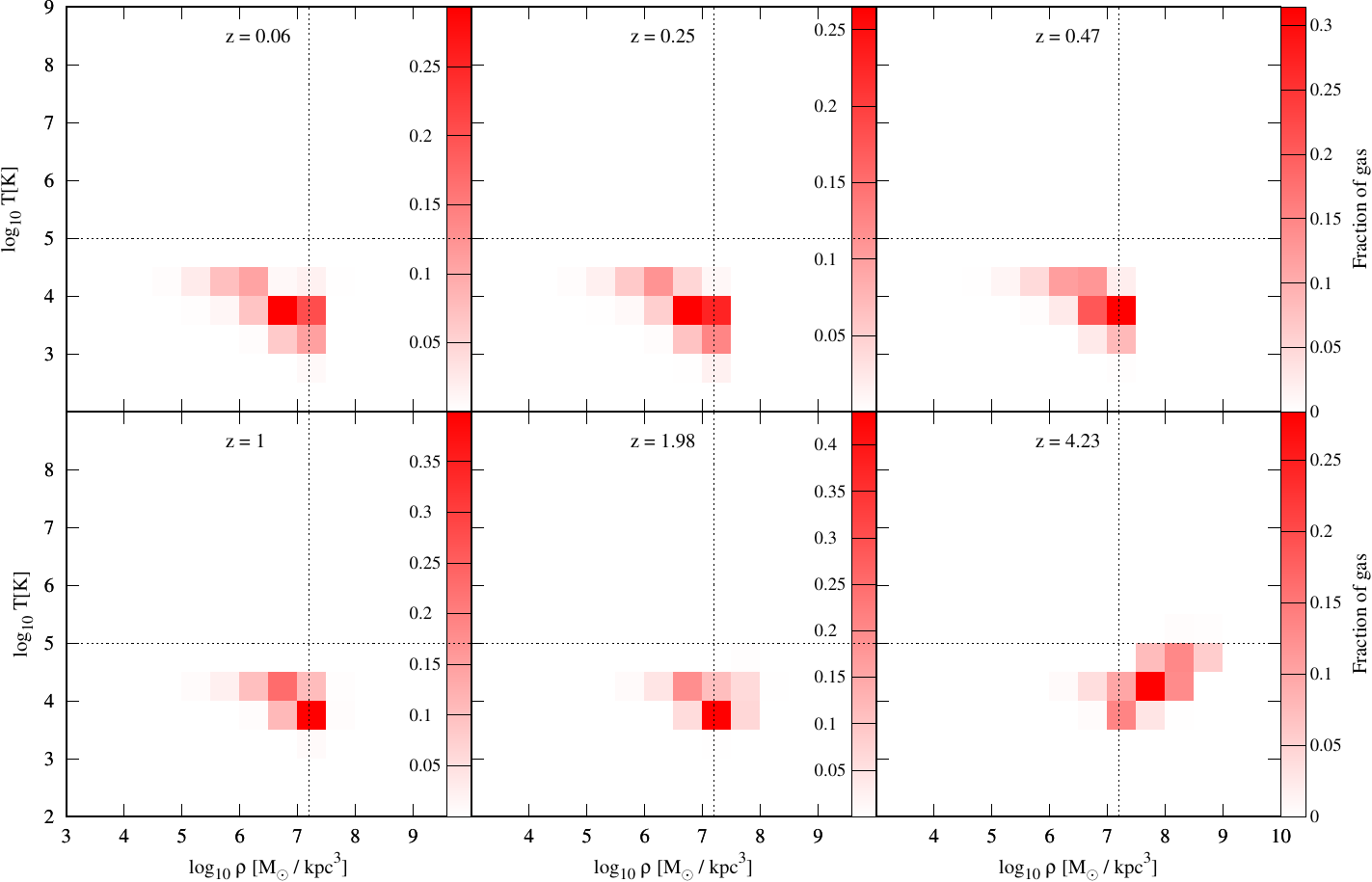}
   \includegraphics[width=0.9\textwidth,clip=true]{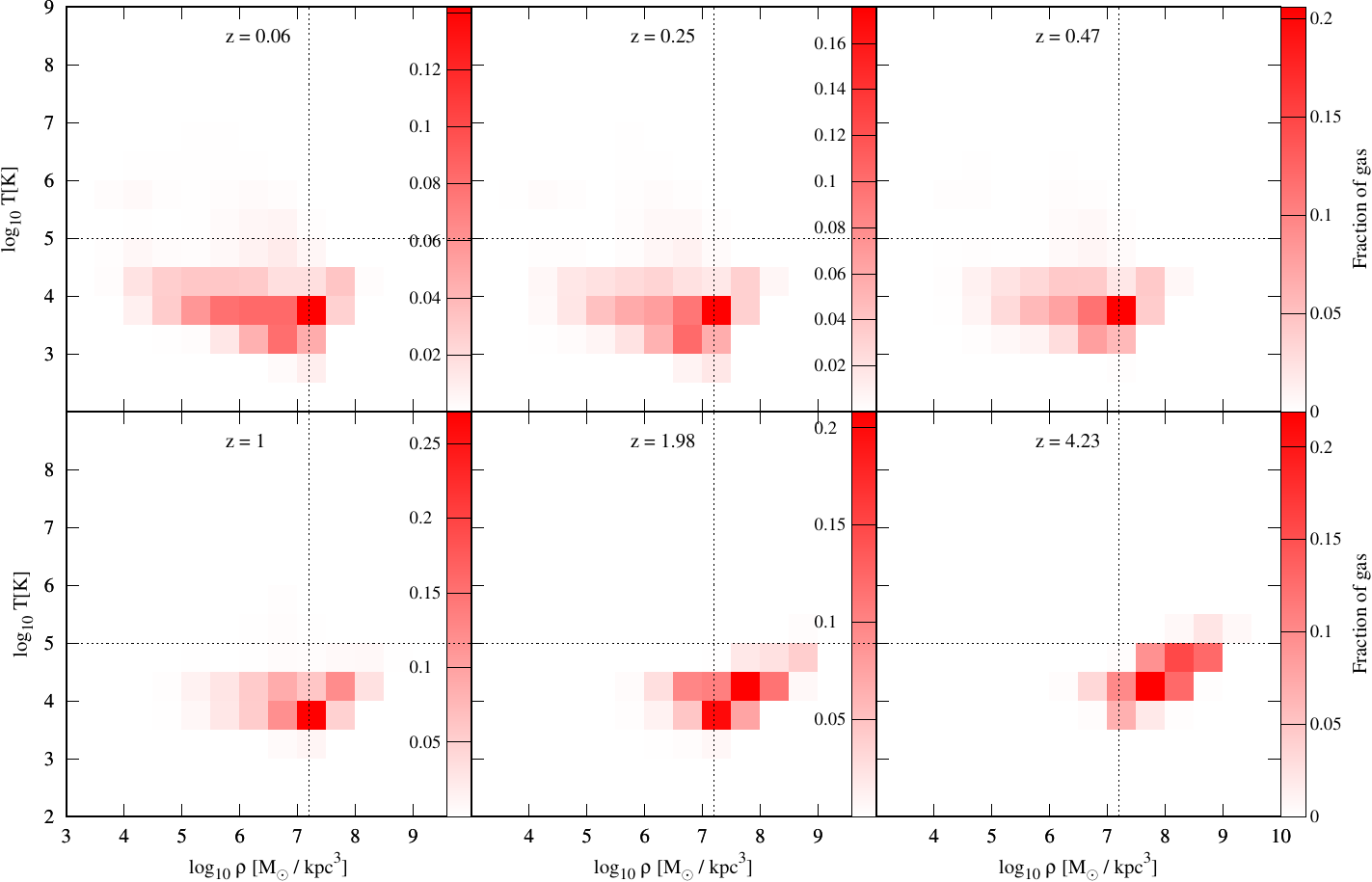}
 \caption{\small{The evolution of the mean temperature vs. density diagram for the galaxies in the lowest (top panels) and intermediate (bottom panels) mass bins of mean values of $log_{10} M_*/M_\odot = 9.28$ and $10.16$, respectively. The intensity of the red color encodes the fraction of gas resolution elements in a grid cell. The vertical dashed line represents the star formation threshold, while the horizontal line shows the temperature cut made for cold gas. }
\label{FigA3} }
\end{center}
\end{figure*}

In Fig. \ref{Fig3} we have shown the evolution of the stacked temperature--density phase diagrams for the gas resolution elements of galaxies in the highest mass bin. Here in Fig. \ref{FigA3}, we show it for two other mass bins, namely the lowest and one intermediate. Again, the intensity of the red color mirrors the fraction of gas in the grid cell. 
Similar to the galaxies in the highest mass bin, at high redshift most of the gas is cold and dense, with a high star formation activity.  
Towards lower redshift there is still a large amount of cold gas, however, less dense, and not reaching the star-forming threshold, any more. 
In contrast to the galaxies in the highest mass bin, at the lowest redshift there is no hot and diffuse gas. This reflects the fact that the galaxies in the lowest mass bin do not have AGNs, which could heat and dissipate the gas. 

\section{Example for Standard Deviations}

\begin{figure}
 \begin{center}
   \includegraphics[width=0.4\textwidth,clip=true]{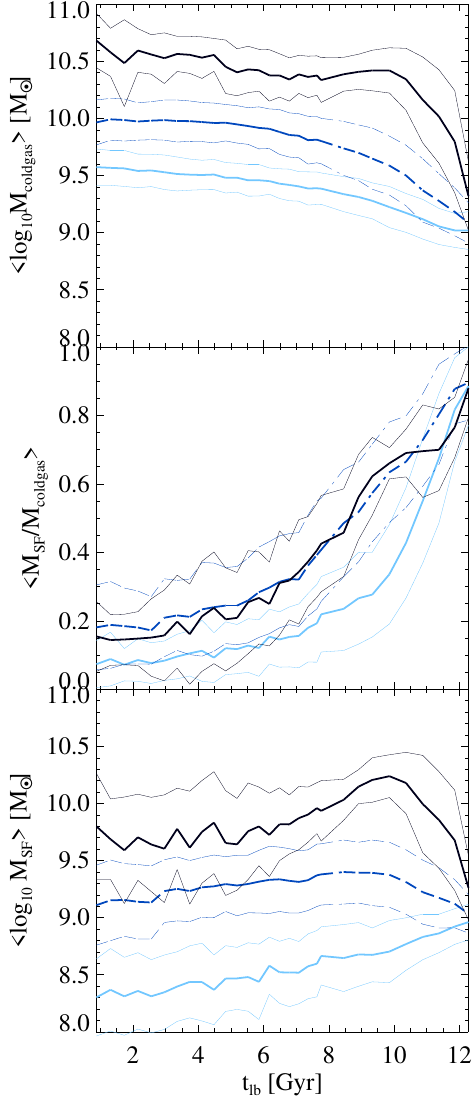}
 \caption{\small{The same as Fig.~\ref{Fig5} but only the lowest, intermediate and highest mass bins are shown in order to illustrate the standard deviation. }
\label{FigA4} }
\end{center}
\end{figure}

Fig. \ref{FigA4} shows the same as Fig. ~\ref{Fig5} but only three mass bins, namely the mass bins of mean values of $log_{10} M_*/M_\odot = 9.28$, $10.16$, and 11.15, respectively. 
In order to indicate the significance of our results we show the lines for $\pm \sigma$. 
We report the values for the minimum and maximum standard deviation over all redshifts separately for the three different mass bins in Table \ref{TabA1}.

\begin{table}
\begin{center}
\begin{tabular}{ |c|c|c| } 
 \hline
 Mass bin $log M_*/M_\odot$ & Min($\sigma(log M_{coldgas}/M_\odot)$) & Max($\sigma(log M_{coldgas}/M_\odot)$) \\
 \hline 
 9.28  & 0.1146 & 0.1803  \\ 
 10.16 & 0.1595 & 0.2776   \\ 
 11.15 & 0.1441 & 0.4478 \\ 
 \hline
 Mass bin $log M_*/M_\odot$ & Min($\sigma(M_{SF}/M_{coldgas})$) & Max($\sigma(M_{SF/}M_{coldgas})$) \\
 \hline
 9.28  & 0.0471 & 0.1890 \\
 10.16 & 0.0986 & 0.1577  \\ 
 11.15 & 0.0456 & 0.1631  \\ 
 \hline
 Mass bin $log M_*/M_\odot$ & Min($\sigma(log M_{SF}/M_\odot)$) & Max($\sigma(log M_{SF}/M_\odot)$) \\
 \hline
 9.28  & 0.1555 & 0.3615 \\
 10.16 & 0.1687 & 0.3536  \\ 
 11.15 & 0.1487 & 0.4944  \\ 
 \hline
\end{tabular}
\caption{\small{The values of the minimum and maximum standard deviations for the quantities of the three mass bins shown in Fig.~\ref{FigA4}. }}
\label{TabA1} 
\end{center}
\end{table}


\section{Global Cold Gas Density}

\begin{figure}
 \begin{center}
   \includegraphics[width=0.5\textwidth,clip=true]{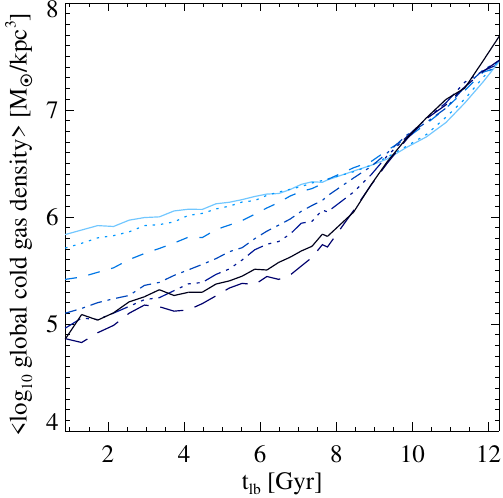}
 \caption{\small{The ``global'' density of the cold gas over time (i.e. the cold gas mass within a sphere divided by the sphere's volume), colored by the final stellar mass as in Fig.~\ref{Fig5}. }
\label{FigA5} }
\end{center}
\end{figure}

In Section \ref{sec:gasprop} we discuss the evolution of the densities of individual gas regions. Here we intent to demonstrate that this is different from the evolution of the  ''global'' density of cold gas in a galaxy. For that purpose we calculate the density of the gas mass in the whole galaxy. 
This is done by using the whole gas mass within a sphere and divide by the volume of this sphere, i.e. $= M_\mathrm{coldgas} / (4/3\cdot \pi \cdot (5 \cdot R_{1/2})^3)$. 
The comparison with observations should be used with caution. 
If a fixed aperture is used for different types or sizes of galaxies, small galaxies can show a low density while large galaxies show a higher density. 
Due to our selection of the gas inside a sphere which is scaling with the stellar half mass radius within the virial radius we mostly avoid this issue.  
In Fig. \ref{FigA5} we can see that it makes a difference compared to the ''local'' gas density (see Fig. \ref{Fig6}). 
The density all over the galaxy is lower than that averaged over individual gas regions.  
This quantity shows implicitly how the mass grows compared to the size. 
For lower mass galaxies the size increases less compared to the gas mass than for higher mass galaxies. 
The most interesting aspect here is that this global density does not tell us if the individual gas regions are dense enough to form stars. i.e. the global density is high for low mass galaxies (light blue) while the local density is lower for them than for the high mass galaxies (dark blue).  

\bibliography{gas.bib}

\begin{thebibliography}{}
\expandafter\ifx\csname natexlab\endcsname\relax\def\natexlab#1{#1}\fi
\providecommand{\url}[1]{\href{#1}{#1}}
\providecommand{\dodoi}[1]{doi:~\href{http://doi.org/#1}{\nolinkurl{#1}}}
\providecommand{\doeprint}[1]{\href{http://ascl.net/#1}{\nolinkurl{http://ascl.net/#1}}}
\providecommand{\doarXiv}[1]{\href{https://arxiv.org/abs/#1}{\nolinkurl{https://arxiv.org/abs/#1}}}

\bibitem[{{Baugh} {et~al.}(2005){Baugh}, {Lacey}, {Frenk}, {Granato}, {Silva},
  {Bressan}, {Benson}, \& {Cole}}]{Baugh2005}
{Baugh}, C.~M., {Lacey}, C.~G., {Frenk}, C.~S., {et~al.} 2005, \mnras, 356,
  1191, \dodoi{10.1111/j.1365-2966.2004.08553.x}

\bibitem[{{Bigiel} {et~al.}(2008){Bigiel}, {Leroy}, {Walter}, {Brinks}, {de
  Blok}, {Madore}, \& {Thornley}}]{Bigiel2008}
{Bigiel}, F., {Leroy}, A., {Walter}, F., {et~al.} 2008, \aj, 136, 2846,
  \dodoi{10.1088/0004-6256/136/6/2846}

\bibitem[{{Bolatto} {et~al.}(2013){Bolatto}, {Wolfire}, \&
  {Leroy}}]{Bolatto2013}
{Bolatto}, A.~D., {Wolfire}, M., \& {Leroy}, A.~K. 2013, \araa, 51, 207,
  \dodoi{10.1146/annurev-astro-082812-140944}

\bibitem[{{Bondi}(1952)}]{Bondi52}
{Bondi}, H. 1952, \mnras, 112, 195

\bibitem[{{Bondi} \& {Hoyle}(1944)}]{Bondi44}
{Bondi}, H., \& {Hoyle}, F. 1944, \mnras, 104, 273

\bibitem[{{Bower} {et~al.}(2006){Bower}, {Benson}, {Malbon}, {Helly}, {Frenk},
  {Baugh}, {Cole}, \& {Lacey}}]{Bower2006}
{Bower}, R.~G., {Benson}, A.~J., {Malbon}, R., {et~al.} 2006, \mnras, 370, 645,
  \dodoi{10.1111/j.1365-2966.2006.10519.x}

\bibitem[{{Cald{\'u}-Primo} {et~al.}(2013){Cald{\'u}-Primo}, {Schruba},
  {Walter}, {Leroy}, {Sandstrom}, {de Blok}, {Ianjamasimanana}, \&
  {Mogotsi}}]{Caldu2013}
{Cald{\'u}-Primo}, A., {Schruba}, A., {Walter}, F., {et~al.} 2013, \aj, 146,
  150, \dodoi{10.1088/0004-6256/146/6/150}

\bibitem[{{Catinella} {et~al.}(2018){Catinella}, {Saintonge}, {Janowiecki},
  {Cortese}, {Dav{\'e}}, {Lemonias}, {Cooper}, {Schiminovich}, {Hummels},
  {Fabello}, {Ger{\'e}b}, {Kilborn}, \& {Wang}}]{Catinella2018}
{Catinella}, B., {Saintonge}, A., {Janowiecki}, S., {et~al.} 2018, \mnras, 476,
  875, \dodoi{10.1093/mnras/sty089}

\bibitem[{{Chabrier}(2003)}]{Chabrier2003}
{Chabrier}, G. 2003, \pasp, 115, 763, \dodoi{10.1086/376392}

\bibitem[{{Chowdhury} {et~al.}(2022){Chowdhury}, {Kanekar}, \&
  {Chengalur}}]{Chowdhury2022}
{Chowdhury}, A., {Kanekar}, N., \& {Chengalur}, J.~N. 2022, \apjl, 935, L5,
  \dodoi{10.3847/2041-8213/ac8150}

\bibitem[{{Crain} {et~al.}(2017){Crain}, {Bah{\'e}}, {Lagos}, {Rahmati},
  {Schaye}, {McCarthy}, {Marasco}, {Bower}, {Schaller}, {Theuns}, \& {van der
  Hulst}}]{Crain2017}
{Crain}, R.~A., {Bah{\'e}}, Y.~M., {Lagos}, C. d.~P., {et~al.} 2017, \mnras,
  464, 4204, \dodoi{10.1093/mnras/stw2586}

\bibitem[{{Dav{\'e}} {et~al.}(2020){Dav{\'e}}, {Crain}, {Stevens}, {Narayanan},
  {Saintonge}, {Catinella}, \& {Cortese}}]{Dave2020}
{Dav{\'e}}, R., {Crain}, R.~A., {Stevens}, A. R.~H., {et~al.} 2020, \mnras,
  497, 146, \dodoi{10.1093/mnras/staa1894}

\bibitem[{{De Lucia} \& {Blaizot}(2007)}]{Delucia2007}
{De Lucia}, G., \& {Blaizot}, J. 2007, \mnras, 375, 2,
  \dodoi{10.1111/j.1365-2966.2006.11287.x}

\bibitem[{{Decarli} {et~al.}(2016){Decarli}, {Walter}, {Aravena}, {Carilli},
  {Bouwens}, {da Cunha}, {Daddi}, {Ivison}, {Popping}, {Riechers}, {Smail},
  {Swinbank}, {Weiss}, {Anguita}, {Assef}, {Bauer}, {Bell}, {Bertoldi},
  {Chapman}, {Colina}, {Cortes}, {Cox}, {Dickinson}, {Elbaz},
  {G{\'o}nzalez-L{\'o}pez}, {Ibar}, {Infante}, {Hodge}, {Karim}, {Le Fevre},
  {Magnelli}, {Neri}, {Oesch}, {Ota}, {Rix}, {Sargent}, {Sheth}, {van der Wel},
  {van der Werf}, \& {Wagg}}]{Decarli2016}
{Decarli}, R., {Walter}, F., {Aravena}, M., {et~al.} 2016, \apj, 833, 69,
  \dodoi{10.3847/1538-4357/833/1/69}

\bibitem[{{Decarli} {et~al.}(2019){Decarli}, {Walter},
  {G{\'o}nzalez-L{\'o}pez}, {Aravena}, {Boogaard}, {Carilli}, {Cox}, {Daddi},
  {Popping}, {Riechers}, {Uzgil}, {Weiss}, {Assef}, {Bacon}, {Bauer},
  {Bertoldi}, {Bouwens}, {Contini}, {Cortes}, {da Cunha}, {D{\'\i}az-Santos},
  {Elbaz}, {Inami}, {Hodge}, {Ivison}, {Le F{\`e}vre}, {Magnelli}, {Novak},
  {Oesch}, {Rix}, {Sargent}, {Smail}, {Swinbank}, {Somerville}, {van der Werf},
  {Wagg}, \& {Wisotzki}}]{Decarli2019}
{Decarli}, R., {Walter}, F., {G{\'o}nzalez-L{\'o}pez}, J., {et~al.} 2019, \apj,
  882, 138, \dodoi{10.3847/1538-4357/ab30fe}

\bibitem[{{Di Matteo} {et~al.}(2005){Di Matteo}, {Springel}, \&
  {Hernquist}}]{DiMatteo05}
{Di Matteo}, T., {Springel}, V., \& {Hernquist}, L. 2005, \nat, 433, 604,
  \dodoi{10.1038/nature03335}

\bibitem[{{Dolag} {et~al.}(2017){Dolag}, {Mevius}, \& {Remus}}]{Dolag2017}
{Dolag}, K., {Mevius}, E., \& {Remus}, R.-S. 2017, Galaxies, 5, 35,
  \dodoi{10.3390/galaxies5030035}

\bibitem[{{Fabjan} {et~al.}(2010{\natexlab{a}}){Fabjan}, {Borgani},
  {Tornatore}, {Saro}, {Murante}, \& {Dolag}}]{Fabjan2010}
{Fabjan}, D., {Borgani}, S., {Tornatore}, L., {et~al.} 2010{\natexlab{a}},
  \mnras, 401, 1670, \dodoi{10.1111/j.1365-2966.2009.15794.x}

\bibitem[{{Fabjan} {et~al.}(2010{\natexlab{b}}){Fabjan}, {Borgani},
  {Tornatore}, {Saro}, {Murante}, \& {Dolag}}]{Fabjan10}
---. 2010{\natexlab{b}}, \mnras, 401, 1670,
  \dodoi{10.1111/j.1365-2966.2009.15794.x}

\bibitem[{{Fall}(1983)}]{Fall1983}
{Fall}, S.~M. 1983, in IAU Symposium, Vol. 100, Internal Kinematics and
  Dynamics of Galaxies, ed. E.~{Athanassoula}, 391--398

\bibitem[{{Font} {et~al.}(2008){Font}, {Bower}, {McCarthy}, {Benson}, {Frenk},
  {Helly}, {Lacey}, {Baugh}, \& {Cole}}]{Font2008}
{Font}, A.~S., {Bower}, R.~G., {McCarthy}, I.~G., {et~al.} 2008, \mnras, 389,
  1619, \dodoi{10.1111/j.1365-2966.2008.13698.x}

\bibitem[{{Gao} \& {Solomon}(2004)}]{Gao2004}
{Gao}, Y., \& {Solomon}, P.~M. 2004, \apj, 606, 271, \dodoi{10.1086/382999}

\bibitem[{{Gil de Paz} {et~al.}(2005){Gil de Paz}, {Madore}, {Boissier},
  {Swaters}, {Popescu}, {Tuffs}, {Sheth}, {Kennicutt}, {Bianchi}, {Thilker}, \&
  {Martin}}]{gildepaz:2005}
{Gil de Paz}, A., {Madore}, B.~F., {Boissier}, S., {et~al.} 2005, \apjl, 627,
  L29, \dodoi{10.1086/432054}

\bibitem[{{Goddard} {et~al.}(2017){Goddard}, {Thomas}, {Maraston}, {Westfall},
  {Etherington}, {Riffel}, {Mallmann}, {Zheng}, {Argudo-Fern{\'a}ndez}, {Lian},
  {Bershady}, {Bundy}, {Drory}, {Law}, {Yan}, {Wake}, {Weijmans}, {Bizyaev},
  {Brownstein}, {Lane}, {Maiolino}, {Masters}, {Merrifield}, {Nitschelm},
  {Pan}, {Roman-Lopes}, {Storchi-Bergmann}, \& {Schneider}}]{Goddard2017}
{Goddard}, D., {Thomas}, D., {Maraston}, C., {et~al.} 2017, \mnras, 466, 4731,
  \dodoi{10.1093/mnras/stw3371}

\bibitem[{{Gonz{\'a}lez Delgado} {et~al.}(2015){Gonz{\'a}lez Delgado},
  {Garc{\'\i}a-Benito}, {P{\'e}rez}, {Cid Fernandes}, {de Amorim},
  {Cortijo-Ferrero}, {Lacerda}, {L{\'o}pez Fern{\'a}ndez}, {Vale-Asari},
  {S{\'a}nchez}, {Moll{\'a}}, {Ruiz-Lara}, {S{\'a}nchez-Bl{\'a}zquez},
  {Walcher}, {Alves}, {Aguerri}, {Bekerait{\'e}}, {Bland-Hawthorn}, {Galbany},
  {Gallazzi}, {Husemann}, {Iglesias-P{\'a}ramo}, {Kalinova},
  {L{\'o}pez-S{\'a}nchez}, {Marino}, {M{\'a}rquez}, {Masegosa}, {Mast},
  {M{\'e}ndez-Abreu}, {Mendoza}, {del Olmo}, {P{\'e}rez}, {Quirrenbach}, \&
  {Zibetti}}]{Gonzalez2015}
{Gonz{\'a}lez Delgado}, R.~M., {Garc{\'\i}a-Benito}, R., {P{\'e}rez}, E.,
  {et~al.} 2015, \aap, 581, A103, \dodoi{10.1051/0004-6361/201525938}

\bibitem[{{Hirschmann} {et~al.}(2014){Hirschmann}, {Dolag}, {Saro}, {Bachmann},
  {Borgani}, \& {Burkert}}]{Hirschmann2014}
{Hirschmann}, M., {Dolag}, K., {Saro}, A., {et~al.} 2014, \mnras, 442, 2304,
  \dodoi{10.1093/mnras/stu1023}

\bibitem[{{Hoyle} \& {Lyttleton}(1939)}]{Hoyle39}
{Hoyle}, F., \& {Lyttleton}, R.~A. 1939, Proceedings of the Cambridge
  Philosophical Society, 35, 405, \dodoi{10.1017/S0305004100021150}

\bibitem[{{Kennicutt}(1998)}]{Kennicutt1998}
{Kennicutt}, Robert~C., J. 1998, \apj, 498, 541, \dodoi{10.1086/305588}

\bibitem[{{Komatsu} {et~al.}(2011){Komatsu}, {Smith}, {Dunkley}, {Bennett},
  {Gold}, {Hinshaw}, {Jarosik}, {Larson}, {Nolta}, \& {Page}}]{Komatsu2011}
{Komatsu}, E., {Smith}, K.~M., {Dunkley}, J., {et~al.} 2011, \apjs, 192, 18,
  \dodoi{10.1088/0067-0049/192/2/18}

\bibitem[{{Kudritzki} {et~al.}(2021){Kudritzki}, {Teklu}, {Schulze}, {Remus},
  {Dolag}, {Burkert}, \& {Zahid}}]{Kudritzki2021}
{Kudritzki}, R.-P., {Teklu}, A.~F., {Schulze}, F., {et~al.} 2021, \apj, 910,
  87, \dodoi{10.3847/1538-4357/abe40c}

\bibitem[{{Lagos} {et~al.}(2011){Lagos}, {Baugh}, {Lacey}, {Benson}, {Kim}, \&
  {Power}}]{Lagos2011}
{Lagos}, C. D.~P., {Baugh}, C.~M., {Lacey}, C.~G., {et~al.} 2011, \mnras, 418,
  1649, \dodoi{10.1111/j.1365-2966.2011.19583.x}

\bibitem[{{Lagos} {et~al.}(2015){Lagos}, {Crain}, {Schaye}, {Furlong}, {Frenk},
  {Bower}, {Schaller}, {Theuns}, {Trayford}, {Bah{\'e}}, \& {Dalla
  Vecchia}}]{Lagos2015}
{Lagos}, C. d.~P., {Crain}, R.~A., {Schaye}, J., {et~al.} 2015, \mnras, 452,
  3815, \dodoi{10.1093/mnras/stv1488}

\bibitem[{{Lenki{\'c}} {et~al.}(2020){Lenki{\'c}}, {Bolatto}, {F{\"o}rster
  Schreiber}, {Tacconi}, {Neri}, {Combes}, {Walter}, {Garc{\'\i}a-Burillo},
  {Genzel}, {Lutz}, \& {Cooper}}]{Lenkic2020}
{Lenki{\'c}}, L., {Bolatto}, A.~D., {F{\"o}rster Schreiber}, N.~M., {et~al.}
  2020, \aj, 159, 190, \dodoi{10.3847/1538-3881/ab7458}

\bibitem[{{Maddox} {et~al.}(2015){Maddox}, {Hess}, {Obreschkow}, {Jarvis}, \&
  {Blyth}}]{maddox:2015}
{Maddox}, N., {Hess}, K.~M., {Obreschkow}, D., {Jarvis}, M.~J., \& {Blyth},
  S.~L. 2015, \mnras, 447, 1610, \dodoi{10.1093/mnras/stu2532}

\bibitem[{{Martin} {et~al.}(2022){Martin}, {Bazkiaei}, {Spavone}, {Iodice},
  {Mihos}, {Montes}, {Benavides}, {Brough}, {Carlin}, {Collins}, {Duc},
  {G{\'o}mez}, {Galaz}, {Hern{\'a}ndez-Toledo}, {Jackson}, {Kaviraj}, {Knapen},
  {Mart{\'\i}nez-Lombilla}, {McGee}, {O'Ryan}, {Prole}, {Rich}, {Rom{\'a}n},
  {Shah}, {Starkenburg}, {Watkins}, {Zaritsky}, {Pichon}, {Armus}, {Bianconi},
  {Buitrago}, {Bus{\'a}}, {Davis}, {Demarco}, {Desmons}, {Garc{\'\i}a},
  {Graham}, {Holwerda}, {Hon}, {Khalid}, {Klehammer}, {Klutse}, {Lazar},
  {Nair}, {Noakes-Kettel}, {Rutkowski}, {Saha}, {Sahu}, {Sola},
  {V{\'a}zquez-Mata}, {Vera-Casanova}, \& {Yoon}}]{martin:2022}
{Martin}, G., {Bazkiaei}, A.~E., {Spavone}, M., {et~al.} 2022, \mnras, 513,
  1459, \dodoi{10.1093/mnras/stac1003}

\bibitem[{{Nagesh} {et~al.}(2023){Nagesh}, {Kroupa}, {Banik}, {Famaey},
  {Ghafourian}, {Roshan}, {Thies}, {Zhao}, \& {Wittenburg}}]{Nagesh2023}
{Nagesh}, S.~T., {Kroupa}, P., {Banik}, I., {et~al.} 2023, \mnras, 519, 5128,
  \dodoi{10.1093/mnras/stac3645}

\bibitem[{{Obreschkow} {et~al.}(2009){Obreschkow}, {Heywood}, {Kl{\"o}ckner},
  \& {Rawlings}}]{Obreschkow2009a}
{Obreschkow}, D., {Heywood}, I., {Kl{\"o}ckner}, H.~R., \& {Rawlings}, S. 2009,
  \apj, 702, 1321, \dodoi{10.1088/0004-637X/702/2/1321}

\bibitem[{{Obreschkow} \& {Rawlings}(2009)}]{Obreschkow2009b}
{Obreschkow}, D., \& {Rawlings}, S. 2009, \apjl, 696, L129,
  \dodoi{10.1088/0004-637X/696/2/L129}

\bibitem[{{Padovani} \& {Matteucci}(1993)}]{Padovani1993}
{Padovani}, P., \& {Matteucci}, F. 1993, \apj, 416, 26, \dodoi{10.1086/173212}

\bibitem[{{Pearson} {et~al.}(2018){Pearson}, {Wang}, {Hurley}, {Ma{\l}ek},
  {Buat}, {Burgarella}, {Farrah}, {Oliver}, {Smith}, \& {van der
  Tak}}]{Pearson2018}
{Pearson}, W.~J., {Wang}, L., {Hurley}, P.~D., {et~al.} 2018, \aap, 615, A146,
  \dodoi{10.1051/0004-6361/201832821}

\bibitem[{{Pety} {et~al.}(2013){Pety}, {Schinnerer}, {Leroy}, {Hughes},
  {Meidt}, {Colombo}, {Dumas}, {Garc{\'\i}a-Burillo}, {Schuster}, {Kramer},
  {Dobbs}, \& {Thompson}}]{Pety2013}
{Pety}, J., {Schinnerer}, E., {Leroy}, A.~K., {et~al.} 2013, \apj, 779, 43,
  \dodoi{10.1088/0004-637X/779/1/43}

\bibitem[{{Popping} {et~al.}(2014){Popping}, {Somerville}, \&
  {Trager}}]{Popping2014}
{Popping}, G., {Somerville}, R.~S., \& {Trager}, S.~C. 2014, \mnras, 442, 2398,
  \dodoi{10.1093/mnras/stu991}

\bibitem[{{Popping} {et~al.}(2019){Popping}, {Pillepich}, {Somerville},
  {Decarli}, {Walter}, {Aravena}, {Carilli}, {Cox}, {Nelson}, {Riechers},
  {Weiss}, {Boogaard}, {Bouwens}, {Contini}, {Cortes}, {da Cunha}, {Daddi},
  {D{\'\i}az-Santos}, {Diemer}, {Gonz{\'a}lez-L{\'o}pez}, {Hernquist},
  {Ivison}, {Le F{\`e}vre}, {Marinacci}, {Rix}, {Swinbank}, {Vogelsberger},
  {van der Werf}, {Wagg}, \& {Yung}}]{Popping2019}
{Popping}, G., {Pillepich}, A., {Somerville}, R.~S., {et~al.} 2019, \apj, 882,
  137, \dodoi{10.3847/1538-4357/ab30f2}

\bibitem[{{Power} {et~al.}(2010){Power}, {Baugh}, \& {Lacey}}]{Power2010}
{Power}, C., {Baugh}, C.~M., \& {Lacey}, C.~G. 2010, \mnras, 406, 43,
  \dodoi{10.1111/j.1365-2966.2010.16481.x}

\bibitem[{{Remus} \& {Forbes}(2022)}]{Remus22}
{Remus}, R.-S., \& {Forbes}, D.~A. 2022, \apj, 935, 37,
  \dodoi{10.3847/1538-4357/ac7b30}

\bibitem[{{Riechers} {et~al.}(2019){Riechers}, {Pavesi}, {Sharon}, {Hodge},
  {Decarli}, {Walter}, {Carilli}, {Aravena}, {da Cunha}, {Daddi}, {Dickinson},
  {Smail}, {Capak}, {Ivison}, {Sargent}, {Scoville}, \& {Wagg}}]{Riechers2019}
{Riechers}, D.~A., {Pavesi}, R., {Sharon}, C.~E., {et~al.} 2019, \apj, 872, 7,
  \dodoi{10.3847/1538-4357/aafc27}

\bibitem[{{Romanowsky} \& {Fall}(2012)}]{Romanowsky2012}
{Romanowsky}, A.~J., \& {Fall}, S.~M. 2012, \apjs, 203, 17,
  \dodoi{10.1088/0067-0049/203/2/17}

\bibitem[{{Saintonge} {et~al.}(2017){Saintonge}, {Catinella}, {Tacconi},
  {Kauffmann}, {Genzel}, {Cortese}, {Dav{\'e}}, {Fletcher},
  {Graci{\'a}-Carpio}, {Kramer}, {Heckman}, {Janowiecki}, {Lutz}, {Rosario},
  {Schiminovich}, {Schuster}, {Wang}, {Wuyts}, {Borthakur}, {Lamperti}, \&
  {Roberts-Borsani}}]{Saintonge2017}
{Saintonge}, A., {Catinella}, B., {Tacconi}, L.~J., {et~al.} 2017, \apjs, 233,
  22, \dodoi{10.3847/1538-4365/aa97e0}

\bibitem[{{Santini} {et~al.}(2017){Santini}, {Fontana}, {Castellano}, {Di
  Criscienzo}, {Merlin}, {Amorin}, {Cullen}, {Daddi}, {Dickinson}, {Dunlop},
  {Grazian}, {Lamastra}, {McLure}, {Micha{\l}owski}, {Pentericci}, \&
  {Shu}}]{santini:2017}
{Santini}, P., {Fontana}, A., {Castellano}, M., {et~al.} 2017, \apj, 847, 76,
  \dodoi{10.3847/1538-4357/aa8874}

\bibitem[{{Schulze} {et~al.}(2018){Schulze}, {Remus}, {Dolag}, {Burkert},
  {Emsellem}, \& {van de Ven}}]{Schulze2018}
{Schulze}, F., {Remus}, R.-S., {Dolag}, K., {et~al.} 2018, \mnras, 480, 4636,
  \dodoi{10.1093/mnras/sty2090}

\bibitem[{{Scoville} {et~al.}(2017){Scoville}, {Lee}, {Vanden Bout},
  {Diaz-Santos}, {Sanders}, {Darvish}, {Bongiorno}, {Casey}, {Murchikova},
  {Koda}, {Capak}, {Vlahakis}, {Ilbert}, {Sheth}, {Morokuma-Matsui}, {Ivison},
  {Aussel}, {Laigle}, {McCracken}, {Armus}, {Pope}, {Toft}, \&
  {Masters}}]{Scoville2017}
{Scoville}, N., {Lee}, N., {Vanden Bout}, P., {et~al.} 2017, \apj, 837, 150,
  \dodoi{10.3847/1538-4357/aa61a0}

\bibitem[{{Shetty} {et~al.}(2014){Shetty}, {Kelly}, {Rahman}, {Bigiel},
  {Bolatto}, {Clark}, {Klessen}, \& {Konstandin}}]{Shetty2014}
{Shetty}, R., {Kelly}, B.~C., {Rahman}, N., {et~al.} 2014, \mnras, 437, L61,
  \dodoi{10.1093/mnrasl/slt156}

\bibitem[{{Sijacki} {et~al.}(2007){Sijacki}, {Springel}, {Di Matteo}, \&
  {Hernquist}}]{Sijacki07}
{Sijacki}, D., {Springel}, V., {Di Matteo}, T., \& {Hernquist}, L. 2007,
  \mnras, 380, 877, \dodoi{10.1111/j.1365-2966.2007.12153.x}

\bibitem[{{Somerville} \& {Dav{\'e}}(2015)}]{Somerville2015}
{Somerville}, R.~S., \& {Dav{\'e}}, R. 2015, \araa, 53, 51,
  \dodoi{10.1146/annurev-astro-082812-140951}

\bibitem[{{Springel} {et~al.}(2005){Springel}, {Di Matteo}, \&
  {Hernquist}}]{Springel05a}
{Springel}, V., {Di Matteo}, T., \& {Hernquist}, L. 2005, \mnras, 361, 776,
  \dodoi{10.1111/j.1365-2966.2005.09238.x}

\bibitem[{{Springel} \& {Hernquist}(2003)}]{Springel2003}
{Springel}, V., \& {Hernquist}, L. 2003, \mnras, 339, 289,
  \dodoi{10.1046/j.1365-8711.2003.06206.x}

\bibitem[{{Steinborn} {et~al.}(2015){Steinborn}, {Dolag}, {Hirschmann},
  {Prieto}, \& {Remus}}]{Steinborn2015}
{Steinborn}, L.~K., {Dolag}, K., {Hirschmann}, M., {Prieto}, M.~A., \& {Remus},
  R.-S. 2015, \mnras, 448, 1504, \dodoi{10.1093/mnras/stv072}

\bibitem[{{Tacconi} {et~al.}(2020){Tacconi}, {Genzel}, \&
  {Sternberg}}]{Tacconi2020}
{Tacconi}, L.~J., {Genzel}, R., \& {Sternberg}, A. 2020, arXiv e-prints,
  arXiv:2003.06245.
\newblock \doarXiv{2003.06245}

\bibitem[{{Teklu} {et~al.}(2015){Teklu}, {Remus}, {Dolag}, {Beck}, {Burkert},
  {Schmidt}, {Schulze}, \& {Steinborn}}]{Teklu2015}
{Teklu}, A.~F., {Remus}, R.-S., {Dolag}, K., {et~al.} 2015, \apj, 812, 29,
  \dodoi{10.1088/0004-637X/812/1/29}

\bibitem[{{Teklu} {et~al.}(2017){Teklu}, {Remus}, {Dolag}, \&
  {Burkert}}]{Teklu2017}
{Teklu}, A.~F., {Remus}, R.-S., {Dolag}, K., \& {Burkert}, A. 2017, \mnras,
  472, 4769, \dodoi{10.1093/mnras/stx2303}

\bibitem[{{Thielemann} {et~al.}(2003){Thielemann}, {Argast}, {Brachwitz},
  {Hix}, {H{\"o}flich}, {Liebend{\"o}rfer}, {Martinez-Pinedo}, {Mezzacappa},
  {Nomoto}, \& {Panov}}]{Thielemann2003}
{Thielemann}, F.~K., {Argast}, D., {Brachwitz}, F., {et~al.} 2003, in From
  Twilight to Highlight: The Physics of Supernovae, ed. W.~{Hillebrandt} \&
  B.~{Leibundgut}, 331

\bibitem[{{Thilker} {et~al.}(2005){Thilker}, {Bianchi}, {Boissier}, {Gil de
  Paz}, {Madore}, {Martin}, {Meurer}, {Neff}, {Rich}, {Schiminovich},
  {Seibert}, {Wyder}, {Barlow}, {Byun}, {Donas}, {Forster}, {Friedman},
  {Heckman}, {Jelinsky}, {Lee}, {Malina}, {Milliard}, {Morrissey}, {Siegmund},
  {Small}, {Szalay}, \& {Welsh}}]{thilker:2005}
{Thilker}, D.~A., {Bianchi}, L., {Boissier}, S., {et~al.} 2005, \apjl, 619,
  L79, \dodoi{10.1086/425251}

\bibitem[{{Tornatore} {et~al.}(2007){Tornatore}, {Borgani}, {Dolag}, \&
  {Matteucci}}]{Tornatore2007}
{Tornatore}, L., {Borgani}, S., {Dolag}, K., \& {Matteucci}, F. 2007, \mnras,
  382, 1050, \dodoi{10.1111/j.1365-2966.2007.12070.x}

\bibitem[{{Valentini} {et~al.}(2019){Valentini}, {Borgani}, {Bressan},
  {Murante}, {Tornatore}, \& {Monaco}}]{Valentini2019}
{Valentini}, M., {Borgani}, S., {Bressan}, A., {et~al.} 2019, \mnras, 485,
  1384, \dodoi{10.1093/mnras/stz492}

\bibitem[{{Valentini} {et~al.}(2022){Valentini}, {Dolag}, {Borgani}, {Murante},
  {Maio}, {Tornatore}, {Granato}, {Ragone-Figueroa}, {Burkert}, {Ragagnin}, \&
  {Rasia}}]{Valentini2022}
{Valentini}, M., {Dolag}, K., {Borgani}, S., {et~al.} 2022, \mnras,
  \dodoi{10.1093/mnras/stac2110}

\bibitem[{{van den Hoek} \& {Groenewegen}(1997)}]{Hoek1997}
{van den Hoek}, L.~B., \& {Groenewegen}, M.~A.~T. 1997, \aaps, 123, 305,
  \dodoi{10.1051/aas:1997162}

\bibitem[{{Walter} {et~al.}(2014){Walter}, {Decarli}, {Sargent}, {Carilli},
  {Dickinson}, {Riechers}, {Ellis}, {Stark}, {Weiner}, {Aravena}, {Bell},
  {Bertoldi}, {Cox}, {Da Cunha}, {Daddi}, {Downes}, {Lentati}, {Maiolino},
  {Menten}, {Neri}, {Rix}, \& {Weiss}}]{Walter2014}
{Walter}, F., {Decarli}, R., {Sargent}, M., {et~al.} 2014, \apj, 782, 79,
  \dodoi{10.1088/0004-637X/782/2/79}

\bibitem[{{Wiersma} {et~al.}(2009){Wiersma}, {Schaye}, \&
  {Smith}}]{Wiersma2009}
{Wiersma}, R. P.~C., {Schaye}, J., \& {Smith}, B.~D. 2009, \mnras, 393, 99,
  \dodoi{10.1111/j.1365-2966.2008.14191.x}

\bibitem[{{Woosley} \& {Weaver}(1995)}]{Woosley1995}
{Woosley}, S.~E., \& {Weaver}, T.~A. 1995, \apjs, 101, 181,
  \dodoi{10.1086/192237}

\end{thebibliography}

\end{document}